\documentclass[journal,twoside]{IEEEtran}

\usepackage{chemarrow}
\usepackage{CJK}
\usepackage{fancybox}
\usepackage{times}
\usepackage{amsmath}
\usepackage{amssymb}
\usepackage[dvips]{graphicx}
\usepackage{subfigure}
\usepackage{bm}
\usepackage{cite}
\usepackage{array}
\usepackage[linesnumbered,ruled]{algorithm2e}

\usepackage{algorithmic}

\newtheorem{proposition}{{Proposition}}

\begin{document}



\title{Pricing-based Distributed Downlink Beamforming  in Multi-Cell OFDMA
Networks}

\author{ Weiqiang~Xu,~\IEEEmembership{Member,~IEEE,}
        Xiaodong~Wang,~\IEEEmembership{Fellow,~IEEE}

\thanks{Manuscript received October 30, 2010; accepted August 29, 2011. This work was  supported in part by the National   Science Foundation of China (NSFC) under grants 60702081, 60873020, 61002016, by the
Joint Research Fund for Overseas Chinese, Hong Kong and Macao Young Scholars under grant 61028001,  by the Key Project of Chinese Ministry of Education under grant 212066,
and by the Zhejiang Provincial Science Foundation  under grants Z1080702, Y1090980, Y12F020196.}
\thanks{W. Xu
is with School of Information Science \& Technology, Zhejiang Sci-Tech University, Hangzhou, 310018, P. R. China. (email: wqxu@zstu.edu.cn).}
\thanks{X. Wang is with  the Department of Electrical Engineering, Columbia University, New York,
NY, 10027, USA.  (e-mail: wangx@ee.columbia.edu).}
}


\maketitle              

\begin{abstract}
We address the problem of downlink beamforming for mitigating the co-channel interference
in multi-cell OFDMA networks.
Based on the network utility maximization  framework,
we formulate the problem as a non-convex optimization problem subject to the per-cell power constraints, in which a general utility function of SINR is used to characterize the network performance.
Some classical utility functions, such as the proportional fairness utility,
the weighted sum-rate utility and the {$\alpha$}-fairness utility, are subsumed
as special cases of our formulation.
To solve the problem in a distributed fashion, we devise an algorithm based on the
non-cooperative  game with pricing mechanism.
We give a sufficient condition for the convergence of the algorithm to the Nash
equilibrium (NE), and
 analyze the information exchange overhead among the base stations.
Moreover,  to speed up the optimization of the beam-vectors at each cell,
we derive an efficient algorithm to solve for the KKT conditions   at each cell.
We provide extensive simulation results to demonstrate that the proposed
distributed multi-cell beamforming algorithm  converges to an NE point in just a few
iterations with low information exchange overhead.
Moreover, it provides   significant performance gains,  especially
under the strong interference scenario, in comparison with several existing
multi-cell interference mitigation schemes,
such as the distributed interference alignment method.


\end{abstract}


\section{Introduction}
\label{sec:introduction}

\IEEEPARstart{I}n multi-cell wireless networks, besides the intra-cell interference caused by
spatial multiplexing within each cell, another impediment arises from inter-cell interference
due to the ever-shrinking cell sizes. Alleviating the effects of inter-cell interference requires the
base stations (BSs) to adjust their transmission schemes collectively.
In fact, inter-cell interference mitigation has been identified as a
key issue for future
wireless  networks. In particular, for downlink transmissions,
if the inter-cell interference is mitigated
via coordinated processing across multiple  BSs,
significant performance gains can be  possibly obtained,
especially for the users at the cell edges.
Therefore, recently, there has been a rapidly growing interest in shifting the design paradigm from the conventional single-cell to the cooperative multi-cell networks \cite{Gesbert2010JSAC}.
Various methods, such as \cite{Zhang :2004},\cite{Ng:2008},\cite{Somekh:2007},
have been proposed to provide network-wide,
macroscopic cooperation among different BSs.
In these studies, it is assumed that the BSs in a multi-cell network are connected via backhaul links to a central processing
unit, which has the global knowledge of the transmitted data from
all the users in the network and the downlink channels from each BS to all the users. Such a fully coordinated
case is sometimes referred to as networked MIMO.
However, for large and dense networks, networked MIMO obviously incurs a substantial
infrastructural and computational
overhead, which increases the system costs and hinders the practical implementations.
This motivates the problem of constrained cooperation,
taking into account many practical factors, e.g., limited backhaul capacity\cite{Shamai:2007},
local cooperation\cite{Simeone:2009}, processing complexity and delay \cite{Tamaki:2007},
imperfect channel state information (CSI) \cite{Marsch:2009}\cite{MAwad2010}, and feedback errors\cite{Papadogiannis:2009}.

On the other hand, future cellular networks are envisioned to be
 distributed systems with autonomous and self-coordinated cells.
 Each BS can make independent and rational  decisions in a decentralized manner, with limited
information exchange with the neighboring BSs.
This motivates the study of distributed multi-cell interference mitigation,
which requires only the local
and neighboring CSI at each BS,
without the need of a central controller, and is therefore much easier to implement.
Based on a generalization of uplink-downlink duality to the
multi-cell setting,  an
iterative  algorithm is proposed in \cite{Dahrouj:2010A} to optimally solve
the multicell downlink beamforming problem
for minimizing either the total weighted transmit power or the
maximum per-antenna power subject to the SINR constraints.
An alternative to the transmit power minimization problem
  is  the rate maximization problem subject to the power constraints, which is in general non-convex.
An approach based on the concept of virtual SINR is proposed in
\cite{ Zakhour:2010D}.
In
\cite{venturino:2010}   an  iterative algorithm is developed for solving the KKT conditions of the
weighted sum-rate maximization problem subject to per-cell power constraints.
However, the proof of  convergence is still an open problem.
 Other related works include  \cite{Zhang}, which
 explores the relationship between the MISO interference channel and the cognitive radio MISO
 channel to devise rate-optimal strategies for decentralized multi-cell cooperative beamforming.

Game theory provides a systematic mathematical framework for the study of competition
 and cooperation   among intelligent and rational
decision makers.
There has been a significant amount of recent research that applies game
theory to resource allocation problems in wireless networks\cite{Jorswieck2009Magazine}-\cite{ZhuHan}.In general, game models can be
classified into two main categories: non-cooperative and cooperative games.
Although non-cooperative game is a useful tool to devise totally distributed algorithms,
 the Nash equilibrium (NE) of the non-cooperative
game may suffer a significant performance degradation compared with the optimal
centralized solution. On the other hand, the cooperative game approach offers
performance gain over the non-cooperative
game, but it requires extensive message exchanges among all players, which implies a large
communication overhead and poor scalability when applied to large networks.
The pricing mechanism
is another alternative to overcome the inefficiency of the non-cooperative game approach.
In this paper, we develop a provably convergent distributed multi-cell beamforming
technique based on the pricing-based non-cooperative game.

In particular,
we formulate the problem as a general network utility maximization problem subject to per-cell power constraints,
which is a non-convex problem.
Examples of the utility functions include the weighted sum-rate utility, the proportional fairness utility,
and the $\alpha$-fairness utility \cite{Mo:2000},
etc.
We treat each cell in the multi-cell network as a player, and design an efficient distributed pricing mechanism to
optimize the network performance through coordination among the players.
We give a sufficient condition for the  convergence of the proposed distributed multi-cell beamforming algorithm.
Moreover,
we  derive an efficient algorithm based on the dual decomposition technique for solving the KKT conditions of the   downlink beamforming problem at each BS.
The proposed technique can converge rapidly to the NE point  with a low  information exchange overhead among the BSs.
It  provides significant
performance gains, especially
in the strong interference scenario, in comparison with several existing approaches,
including the recently proposed distributed interference alignment method \cite{Gomadam2009}.

The remainder of this paper is organized as follows. In Section II
we introduce  the system model and the problem formulation.
In Section III we develop the pricing-based multicell distributed downlink beamforming
technique. In Section IV, we derive the beamforming optimization algorithm at each cell
 based on dual decomposition. Simulation results are given in Section V. Finally
Section VI concludes the paper.


\section{System Model and Problem Formation}
\label{sec:system and model}

We consider a downlink multi-cellular network where a set of BSs
$\mathcal{M}=\{1,2,\ldots,M\}$ simultaneously transmit on the
orthogonal sub-channels\footnote{The sub-channel refers to a
logical collection of physical sub-carriers, which is regarded as
the minimum granularity of the radio resource allocation unit in
this paper.} $\mathcal{N}=\{1,2,\ldots,N\}$ during each scheduling
interval. Each BS $m \in \mathcal{M}$ is equipped with $T$ transmit
antennas and  space-division multiple-access (SDMA)  is employed to
serve multiple single-antenna mobile users on each sub-channel.
 Let $\mathcal{B}_{m}^{(n)}$ be the set
of users scheduled by BS $m \in \mathcal{M}$ on sub-channel $n  \in
\mathcal{N}$. For simplicity and without loss of generality, we
assume that $|\mathcal{B}_{m}^{(n)}|=Q, \forall m, \forall n$. We further
assume that each user is served by only one BS.

For data transmission, BS $m$ on sub-channel $n$ transmits  complex
symbols $b_{m,k}^{(n)}\in \mathbb{C}$ through $T$ transmit antennas
using a beam-vector $\mathbf{w}_{m,k}^{(n)}\in \mathbb{C}^{T}$ to
user $k \in \mathcal{B}_{m}^{(n)}$. We assume that
$\mathbb{E}\{|b_{m,k}^{^{(n)}}|^{2}\} = 1$, and
$\mathbb{E}\{b_{m_{1},k_{1}}^{(n_{1})}b_{m_{2},k_{2}}^{(n_{2})}\} =
0$, for $(n_{1},m_{1}, k_{1})\neq(n_2,m_2, k_2)$, where
$\mathbb{E}\{\cdot\}$ is the expectation operator. Then after
normalized by the noise standard deviation, the received signal by
user $k \in \mathcal{B}_{m}^{(n)}$ on sub-channel $n$ can be written
as
\begin{equation}\label{Equ1}
\begin{split}
y_{m,k}^{(n)}
 &=\underbrace{\vec{\mathbf{h}}_{m,k}^{(n)}\mathbf{w}_{m,k}^{(n)}b_{m,k}^{(n)}}_{\mathrm{useful\;signal}}
+\underbrace{\sum_{\substack{k' \in \mathcal{B}_{m}^{(n)}
 \backslash k}}{\vec{\mathbf{h}}_{m,k}^{(n)}\mathbf{w}_{m,k'}^{(n)}b_{m,k'}^{(n)}}}_{\mathrm{in-cell\;co-channel\;interference}}\\
 &\;\;\;\;+\underbrace{\sum_{\substack{j \in \mathcal{M}
\backslash m}} \sum_{u\in \mathcal{B}_{j}^{(n)}}{\vec{\mathbf{h}}_{j,k}^{(n)}\mathbf{w}_{j,u}^{(n)}b_{j,u}^{(n)}}}_{\mathrm{out-cell\;co-channel\;
 interference}}+\underbrace{z_{m,k}^{(n)}}_{\mathrm{noise}},
 \end{split}
\end{equation}
where $\mathbf{h}_{m,k}^{(n)}\in \mathbb{C}^{T}$ is the complex
channel vector between BS $m$ and user $k \in \mathcal{B}_{m}^{(n)}$
on sub-channel $n$, $z_{m, k}^{(n)} \sim {\cal N}_{\mathbb{C}}(0,1)$
denotes the  circularly symmetric complex Gaussian noise sample, and
$\vec{\cdot}$ is the Hermitian transpose operator.

The SINR for user $k \in \mathcal{B}_{m}^{(n)}$ on sub-channel $n$
can then be expressed as\footnote{We drop the explicit dependency of $\Gamma_{m,k}^{(n)}$ and  $\mathcal{I}_{m,k}^{(n)}$ on $\mathbf{W^{(n)}}$.}
\begin{equation}\label{Equ2}
 \Gamma_{m,k}^{(n)}=\frac{\left|  \vec{\mathbf{h}}_{m,k}^{(n)}\mathbf{w}_{m,k}^{(n)}\right| ^{2}}
 {1+\mathcal{I}_{m,k}^{(n)}},
\end{equation}
where $\mathbf{W}^{(n)}=\{\mathbf{w}_{m,k}^{(n)}, k \in
\mathcal{B}_m^{(n)}, m \in \mathcal{M}\}, n  \in \mathcal{N}$, and
\begin{equation}\label{Interference}
\mathcal{I}_{m,k}^{(n)}
=\underbrace{\sum_{\substack{k' \in \mathcal{B}_m^{(n)}
 \backslash k}}{\left|  \vec{\mathbf{h}}_{m,k}^{(n)}\mathbf{w}_{m,k'}^{(n)}\right| ^{2}}}_{\mathcal{I}_{m,k}^{(n),\texttt{in}}} +\underbrace{\sum_{\substack{j \in \mathcal{M}
 \backslash m}}\underbrace{\sum_{u \in \mathcal{B}_{j}^{(n)}}\left|  \vec{\mathbf{h}}_{j,k}^{(n)}\mathbf{w}_{j,u}^{(n)}
 \right|
 ^{2}}_{\mathcal{I}_{m,k}^{(n),\texttt{out},j}}}_{\mathcal{I}_{m,k}^{(n),\texttt{out}}},
\end{equation}
where the terms $\mathcal{I}^{(n),\texttt{in}}_{m,k}$ and
$\mathcal{I}^{(n),\texttt{out}}_{m,k}$ account for the in-cell and
out-cell interference, respectively.

Now, we consider the following general linear beamforming
optimization problem where we wish  to maximize a network-wide
utility function across all users of all coordinated BSs and all
sub-channels, by choosing the set of beam-vectors $\mathbf{W}=\{
\mathbf{W}^{(n)}, n \in {\cal N}\}$, subject to the
per-base-station power constraints:
\begin{equation}\label{Equ4}
\begin{split}
 \max_{\mathbf{W}}  & \ \ \ U_{{\rm network}}(\mathbf{W})=\sum_{m \in \mathcal{M}}\sum_{n \in \mathcal{N}} \sum_{k \in \mathcal{B}_m^{(n)}} U_{m,k}^{(n)}\Big(\Gamma_{m,k}^{(n)}
 \Big),\\
  \mathrm{s.t.} & \ \ \ \sum_{n \in \mathcal{N}}\sum_{k \in
\mathcal{B}_{m}^{(n)}}\vec{\mathbf{w}}_{m,k}^{(n)}\mathbf{w}_{m,k}^{(n)}
\leq  {P}_m, \; \forall m \in \mathcal{M},
\end{split}
\end{equation}
where ${P}_m$ is the total transmit power at BS $m$. We assume that
the above optimization problem has a set of feasible solutions,
which can be facilitated  through some form of admission control
  or/and scheduling strategies.

In the above formulation, each user $k \in \mathcal{B}_m^{(n)}$ is
assigned a utility function $U_{m,k}^{(n)}(\Gamma_{m,k}^{(n)})$,
which is
assumed to be a monotonically nondecreasing, concave and twice
differentiable function of the received SINR $\Gamma_{m,k}^{(n)}$. Typical
utility functions include the following:
\begin{itemize}
\item Proportional fairness utility \cite{Mo:2000}: $U (\Gamma ) = \log(\Gamma )$;
\item Rate utility: $U (\Gamma ) =\log(1+\Gamma )$;
\item $\alpha$-fairness utility \cite{Mo:2000}:
$
U (\Gamma ) =
   (1-\alpha)^{-1}(\Gamma )^{1-\alpha}, \ \ \alpha \neq 1.
$
\end{itemize}

Note that the constraint set in (\ref{Equ4}) is convex. However, due
to the SINR expression (\ref{Equ2}), even though the utility function
$U_{m,k}^{(n)}(\Gamma_{m,k}^{(n)})$ is concave in terms of the SINR $\Gamma_{m,k}^{(n)}$, it is in
general nonconcave in terms of the set of beam-vectors
$\mathbf{W}^{(n)}$. Numerically finding the global optimal solution to the
optimization problem (\ref{Equ4}) is known to be a difficult problem.
Our
objective is to develop a distributed solution to (\ref{Equ4}) where
each BS updates its beam-vectors locally; and with the aid of
limited information exchange among the BSs, some form of optimality
can be achieved. To that end, we resort to the game theoretical tool
of pricing mechanism.


\section{Pricing Mechanism and Distributed Algorithm for Non-Cooperative Beamforming Game}
An extreme example of distributed beamforming scheme  is for each BS
to independently update its own beam-vectors without considering the
actions of other BSs.  However, such a pure non-cooperative approach
  may result in non-convergence or some undesirable Nash equilibrium
  (NE) with low   individual as well as system-wise performance \cite{Larsson:2008}.
For instance, it is shown in  \cite {Jorswieck:2008}   that for a
two-user MISO system,  the NE point achieved through  the pure
non-cooperative game over all possible choices of beams is far away
from the Pareto boundary of the achievable rate region.

The pricing mechanism \cite{Saraydar:2002,Huang:2006,Schmidt:2009} has been employed
as an effective means to stimulate cooperation among players, and to
guide the players' behaviors toward a more efficient NE that
improves the system performance, by introducing a certain degree of
coordination in a non-cooperative game.
 In this section, we propose a pricing mechanism for the
 non-cooperative multicell beamforming game and the corresponding
 distributed beamforming algorithm. We then prove the convergence of
 this algorithm.
 Finally we analyze the information exchange
 overhead among the BSs.

\subsection{Pricing Mechanism}
We  model the pricing-based non-cooperative multicell beamforming game
  as
$$\mathcal{G}=\{\mathcal{M},\{\mathcal{W}_m\}_{m \in
\mathcal{M}},\{\bar{U}_m\}_{m \in \mathcal{M}}\},$$ where the
elements   are
\begin{itemize}
  \item Player set: $\mathcal{M}=\{1,2,\ldots,M\}$, i.e., the set of BSs.
  \item Strategy set: $\{\mathcal{W}_1,\ldots,\mathcal{W}_M\}$, where the strategy set of player
   (BS) $m$ is the following
\begin{equation}\label{Equ5}
\begin{split}
\mathcal{W}_m &=\Big\{\mathbf{w}_{m,k}^{(n)} \in \mathbb{C}^T, k \in
\mathcal{B}_m^{(n)}, n\in \mathcal{N}:\\
 &\;\;\;\;\;\;\sum_{n \in \mathcal{N}}\sum_{k \in \mathcal{B}_m^{(n)}}\vec{\mathbf{w}}_{m,k}^{(n)}\mathbf{w}_{m,k}^{(n)}\leq {P}_m\Big\}.\\
 \end{split}
\end{equation}
\item Payoff functions set: $\{\bar{U}_1, \ldots,\bar{U}_M \}$,
with
\begin{equation}\label{Equ7 8}
\begin{split}
\;\;\;\;\;\;\;\;\; \bar{U}_m(\mathbf{W}_m,\mathbf{W}_{-m})\\
&\hspace{-3.1cm}=\sum_{n \in
\mathcal{N}}\sum_{k \in \mathcal{B}_m^{(n)}}{U_{m,k}^{(n)}
(\Gamma_{m,k}^{(n)})}-C(\mathbf{W}_m,\mathbf{W}_{-m}),
\end{split}
\end{equation}
where
$\mathbf{W}_{m}=\{\mathbf{w}_{m,k}^{(n)}, k \in \mathcal{B}_m^{(n)},
n\in \mathcal{N}\}$ and
$\mathbf{W}_{-m}=\{\mathbf{W}_1,...,\mathbf{W}_{m-1},\mathbf{W}_{m+1},...\mathbf{W}_M\}$
denote the set of beam-vectors of BS $m$, and that of all other BSs,
respectively. $C(\mathbf{W}_m,\mathbf{W}_{-m})$ is a  cost function
associated with a pricing mechanism.
 \end{itemize}


An efficient pricing mechanism should take into account the nature of the
service requirement of each player  and  reflect accurately the cost of
resource consumption for fullfilling each player's requirement.
Inspired
by \cite{Saraydar:2002,Huang:2006,Schmidt:2009},
we will apply the usage-based
pricing mechanism to solve our problem, where the price a player pays for using the resource is proportional
to the amount of resource consumed by the player.

First, we introduce
a quantity called the interference pricing
rate  of  user $k \in \mathcal{B}_m^{(n)}$, which
measures the marginal decrease in utility due to a marginal
increase in interference, given by
\begin{equation}\label{Equ_price}
  \pi_{m,k}^{(n)}\triangleq-\frac{\partial U_{m,k}^{(n)}}{\partial \mathcal{I}_{m,k}^{(n)}}
   =(U_{m,k}^{(n)})'\frac{|\vec{\mathbf{h}}_{m,k}^{(n)}\mathbf{w}_{m,k}^{(n)}|^{2}}
   {(1+\mathcal{I}_{m,k}^{(n)})^{2}},
\end{equation}
where $(U_{m,k}^{(n)})'$ denote the  derivative  of the utility
function  with respect to the SINR  $\Gamma_{m,k}^{(n)}$.
When BS $m$
transmits signal to user $k \in \mathcal{B}_m^{(n)}$ on
sub-channel $n$ using the beam-vector $\mathbf{w}_{m,k}^{(n)}$,
it induces the interference $\left| \vec{\mathbf{h}}_{m,u}^{(n)}
\mathbf{w}_{m,k}^{(n)} \right|^2$ to all other users ${u \in
\mathcal{B}_j^{(n)}}, (j,u) \neq (m,k), j \in \mathcal{M}$. Thus,
under the pricing mechanism, when serving  user $k \in \mathcal{B}_m^{(n)}$,
BS $m$ needs to pay a total cost:
\begin{equation}
\sum_{j \in
\mathcal{M}} {\sum_{u \in \mathcal{B}_j ^{(n)}} {\pi _{j,u}^{^{(n)}} \left| {
{\vec{\mathbf{h}}_{m,u}^{^{(n)}} }  \mathbf{w}_{m,k}^{^{(n)}} }
\right|^2 } }
  = {\vec{\mathbf{w}}_{m,k}^{^{(n)}} } \mathbf{L}_{m,k}^{^{(n)}} \mathbf{w}_{m,k}^{^{(n)}},
\end {equation}

 \begin{figure*}
  \suppressfloats[!b]
  \begin{equation}
\label{LeakMatrix}
 \mathbf{L}_{m,k}^{(n)}\triangleq
\underbrace{\sum_{\substack{k' \in \mathcal{B}_m^{(n)}
 \backslash k}}\pi_{m,k'}^{(n)} \mathbf{h}_{m,k'}^{(n)}\vec{\mathbf{h}}_{m,k'}^{(n)}}_{\mathbf{L}_{m,k}^{(n),\texttt{in}}}
 +\underbrace{\sum_{\substack{j \in \mathcal{M}
\backslash m}} \underbrace{\sum_{u \in
\mathcal{B}_{j}^{(n)}}\pi_{j,u}^{(n)} \mathbf{h}_{m,u}^{(n)}
\vec{\mathbf{h}}_{m,u}^{(n)}}_{\mathbf{L}_{m}^{(n),\texttt{out},j}}}_{\mathbf{L}_{m}^{(n),\texttt{out}}}.
\end{equation}
\hrulefill
\end{figure*}

where $\mathbf{L}_{m,k}^{(n)}$ is defined in (\ref{LeakMatrix}). We called $\mathbf{L}_{m,k}^{(n)}$ as the leakage matrix of user $k
\in \mathcal{B}_m^{(n)}$ on sub-channel $n$, which accounts for the
amount of interference caused by BS $m$ to other co-channel users on
sub-channel $n$ when serving user $k \in \mathcal{B}_m^{(n)}$. Note
that $\mathbf{L}_{m,k}^{(n)}$ is Hermitian symmetric, i.e.,
$\mathbf{L}_{m,k}^{(n)}=\vec{\mathbf{L}}_{m,k}^{(n)}$, since $
\mathbf{h}_{m,k}^{(n)}\vec{\mathbf{h}}_{m,k}^{(n)}$ is Hermitian
symmetric. The terms $\mathbf{L}_{m,k}^{(n),\texttt{in}}$ and
$\mathbf{L}_{m}^{(n),\texttt{out}}$ in (\ref{LeakMatrix})
account for the in-cell and
out-cell leakages, respectively.

Hence summing across all users served by BS $m$ and across all
sub-channels, BS $m$ needs to pay a total cost of
\begin{equation}\label{totalprice2}
C(\mathbf{W}_m,\mathbf{W}_{-m})=\sum_{n \in \mathcal{N}} \sum_{k \in
\mathcal{B}_m^{(n)}}\vec{\mathbf{w}}_{m,k}^{(n)}\mathbf{L}_{m,k}^{(n)}\mathbf{w}_{m,k}^{(n)}.
\end{equation}

Summarizing the discussion above,  in the pricing-based
non-cooperative multicell beamforming game, each BS $m$   solves the
following optimization problem
 \begin{equation}\label{BSSolver}
\begin{split}
 \max_{  \mathbf{W}_m }  &
\ \sum_{n \in \mathcal{N}}
\sum_{k \in \mathcal{B}_m^{(n)}} \Big (U_{m,k}^{(n)}
\Big(\Gamma_{m,k}^{(n)}\Big)-
 \vec{\mathbf{w}}_{m,k}^{(n)}\mathbf{L}_{m,k}^{(n)}\mathbf{w}_{m,k}^{(n)} \Big),\\
   \mathrm{s.t.} & \ \ \   \sum_{n \in \mathcal{N}}\sum_{k \in \mathcal{B}_m^{(n)}} \vec{\mathbf{w}}_{m,k}^{(n)}\mathbf{w}_{m,k}^{(n)}\leq
   P_m.
\end{split}
\end{equation}
Notice that the objective function is still  nonconcave with respect
to the beam-vectors $\mathbf{W}_{m}$ associated with BS $m$;
thus the globally optimal solution to (\ref{BSSolver}) cannot be found.
In Section~\ref{dd.sec} we drive a dual decomposition algorithm for
obtaining the solution to the KKT conditions of (\ref{BSSolver}).

\subsection{Distributed Multicell Beamforming Algorithm}

We propose the following distributed algorithm for implementing the
pricing-based non-cooperative multicell beamforming game.

\begin{algorithm}
\textsf{Initialization}:\\
$\;\;$ Each BS $m$ initializes $
\mathbf{W}_m$ satisfying the
power constraint.\\
\textsf{Repeat}\\
$\;\;$ $ \textsf{{For}} $ $\;$ $ m =1:M$ \\
$\;\;\;\; \;\; \;\;$  BS $m$ obtains a solution $\bar{\mathbf{W}}_m$ to (\ref{BSSolver}) for given\\
$\;\;\;\; \;\; \;\;$  $\mathbf{W}_{-m}$, using Algorithm 2.\\
$\;\;\;\; \;\; \;\;$ \textsf{If} $\bar{U}_m(\bar{\mathbf{W}}_m,\mathbf{W}_{-m}) \geq
  \bar{U}_m(\mathbf{W}_m,\mathbf{W}_{-m})$\\
$\;\;\;\; \;\; \;\;$ \textsf{Then}\\
$\;\;\;\; \;\; \;\; \; \;\;$ $\{$\\
$\;\;\;\; \;\; \;\; \; \;\; \;\;$ BS $m$ updates its beam-vectors as $\bar{\mathbf{W}}_m$.\\
$\;\;\;\; \;\; \;\;\; \;\;\;\;  $ Based on the new beam-vectors $\bar{\mathbf{W}}_m$, BS $m$\\
$\;\;\;\; \;\; \;\;\; \;\;\;\;  $ updates\\
$\;\;\;\; \;\; \;\;\; \;\;\;\;  $ $\{\mathcal{I}_{j,u}^{(n),\texttt{out},m},\;
 u \in \mathcal{B}_{j}^{(n)}, \;j \in \mathcal{M}\backslash m,  \;n \in \mathcal{N}\}$,\\
 $\;\;\;\; \;\; \;\;\; \;\;\;\; $ and $\{\pi_{m,k}^{(n)}, \;
 k \in \mathcal{B}_{m}^{(n)},  \;n \in \mathcal{N}\}$ \\
$\;\;\;\; \;\; \;\;\; \;\;\;\;  $ according to (\ref{Interference}) and  (\ref{Equ_price}) respectively,
and\\
$\;\;\;\; \;\; \;\;\; \;\;\;\;  $ passes them to BSs $  j \in \mathcal{M} \backslash m$.\\
$\;\;\;\; \;\; \;\; \; \;\;$ $\}$\\
$\;\;\;\; \;\; \;\;$ $\textsf{EndIf}$\\
$\;\; $ $\textsf{EndFor}$\\
$\textsf{Until}$ convergence\\
\caption{Distributed multicell beamforming algorithm.
\label{DPSPB}}
\end{algorithm}

We have the following observations on Algorithm \ref{DPSPB}.
\begin{enumerate}
\item Only
one BS updates its beam-vectors at a time,  based on the latest
out-cell interference powers and interference price rates (and
thus the latest out-cell leakage matrices) from every other BS in
the multicell network.  Moreover, after a BS updates its
beam-vectors, the new out-cell interference powers and new
interference price rates are announced timely to every other BS.

\item Only if $\bar{U}_m(\bar{\mathbf{W}}_m,\mathbf{W}_{-m}) \geq
  \bar{U}_m(\mathbf{W}_m,\mathbf{W}_{-m})$ holds, BS $m$ updates its beam-vectors as $\bar{\mathbf{W}}_m$.
  Otherwise, BS $m$ keeps its old beam-vectors.
This method is based on the better response strategy in game theory,  which
refers to an update procedure where the players choose actions that
increase their utilities as opposed to maximizing their utilities in
the best response strategy. Notice that the best response strategy cannot applied
  due to the nonconvexity of  (\ref{BSSolver}).
\end{enumerate}
These features ensure the convergence of the algorithm, as discussed
next.

\subsection{Existence and Convergence of NE}
The Nash equilibrium (NE) is a well-known concept for analyzing a
game. A set of  beam-vectors
$\mathbf{W}^{*}=(\mathbf{W}_1^{*},\ldots,\mathbf{W}_M^{*})$ is an NE
 if, for every BS $m \in \mathcal{M}$,
$\bar{U}_m(\mathbf{W}_m^{*},\mathbf{W}_{-m}^{*})\geq
\bar{U}_m(\mathbf{W}_m,\mathbf{W}_{-m}^{*})$,  $\forall \mathbf{W}_m
\in \mathcal{W}_m$. That is,  given the other BSs' beam-vectors, no
BS can increase its utility unilaterally by changing its own
beam-vectors. For the multicell beamforming game under
consideration,  the existence and convergence of NE is heavily
dependent on the concavity of the utility function $U_{m,k}^{(n)}$. We
first introduce a quantity that measures the relative concavity
of a utility function. Specifically, the coefficient of relative
risk aversion associated with the utility function
$U (\Gamma )$ is defined as
\begin{equation}
    \kappa (\Gamma )=-\frac{\Gamma \cdot
     U (\Gamma ) ''}{ U (\Gamma ) '},
\end{equation}
where $ U (\Gamma ) '$ and $ U (\Gamma ) ''$ denote  the first- and second-order derivatives, respectively.

We have the following result on a sufficient condition for the
convergence of Algorithm 1.
\begin{proposition}
Suppose that the utility function $U_{m,k}^{(n)}$ satisfies
 $$ 0 \leq \kappa_{m,k}^{^{(n)}} \leq 2, \ \
 \forall   k \in \mathcal{B}_{m}^{(n)}, \;
 m \in \mathcal{M}, \; n \in \mathcal{N},$$ then Algorithm 1 converges to an NE point.
\end{proposition}
\noindent {\it Proof:}  See Appendix A. \hfill $\Box$

\medskip
\noindent
{\bf Remark:} \  The condition $0 \leq
\kappa_{m,k}^{^{(n)}} \leq 2 $  can be interpreted as requiring that
the utility function  to be sufficiently concave, but not too concave. If the
utility function is too concave (i.e., $\kappa_{m,k}^{^{(n)}} > 2$), the
updates may be too aggressive to guarantee convergence. Fortunately,
this condition  is satisfied by
most utility functions of interest, as discussed below.
\begin{enumerate}
\item For the proportional fairness utility function
 $U (\Gamma  )= \log_2( \Gamma )$,
its coefficient of relative risk aversion is $\kappa =1$.

\item For the   {$\alpha$}-fairness utility function
$ U(\Gamma) =
\frac{(\Gamma )^{1-\alpha}}{1-\alpha} $  with $\alpha \neq 1$,  we have $\kappa_{m,k}^{(n)}=\alpha$.
Hence, for $0 \leq \alpha \leq 2 \ (\alpha \neq 1)$, we have $0 \leq
\kappa  \leq 2  \ (\kappa  \neq 1)$.

\item For the weighted sum-rate utility
$U(\Gamma) = \omega \log_2 (1 + \theta\Gamma)$, with $0 < \theta \leq 1$, we have the following.

\begin{itemize}
\item $\theta=1$ corresponds to the Shannon rate with
 $0 < \kappa =\frac{\Gamma }{1+\Gamma } < 1 $.

\item  $0<\theta<1$ corresponds to the achievable
rate for some practical modulations,
where $\theta =  - \frac{{\phi_1 }}{{\log \left( {\phi_2 {\rm{BER}}}
\right)}}$, and $\phi_1, \phi_2$ are constants depending on the
modulation and BER is the required bit-error rate.
We have $0 < \kappa =\frac{\theta\Gamma }{1+\theta\Gamma } < 1 $.
\end{itemize}

\end{enumerate}

\subsection{Information Exchange Overhead}

 \begin{figure}[htbp]
 \centering
  \includegraphics[scale=0.6,bb=108 360 434 510]{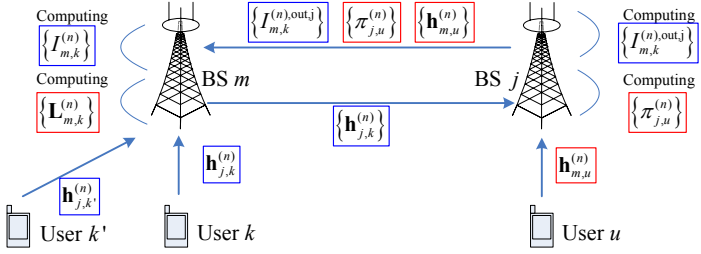}
  \centering
    \caption{Local computation at each BS and the information exchange among the BSs.}
 \label{InformationExchange}
 \end{figure}

In order to compute the SINR $\Gamma_{m,k}^{(n)}$ and the leakage matrix
  ${\bf L}_{m,k}^{(n)}$ in (\ref{BSSolver}), BS $m$ needs to get certain  information from the neighboring BSs. In Fig. \ref{InformationExchange}, we give a graphic illustration of the local computation at each BS and the information exchange among the BSs, which is further elaborated as follows.

BS $m$ can compute $\mathcal{I}_{m,k}^{(n)}$
in (\ref{Equ2})-(\ref{Interference})
for each user $ k \in \mathcal{B}_{m}^{(n)}$ on each
sub-channel $n \in \mathcal{N}$ through the following.
\begin{enumerate}
\item
 Computing the first term $\mathcal{I}_{m,k}^{(n),\texttt{in}}=\sum_{\substack{k' \in \mathcal{B}_m^{(n)}
 \backslash k}}{\left|  \vec{\mathbf{h}}_{m,k}^{(n)}\mathbf{w}_{m,k'}^{(n)}\right| ^{2}}$ in (\ref{Interference}). BS $m$ only needs the knowledge of the direct downlink channels $\mathbf{h}_{m,k}^{(n)}$ and of
the local beam-vectors $\{\mathbf{w}_{m,k'}^{(n)}, k' \in \mathcal{B}_{m}^{(n)}\backslash k\}$ computed at the previous
iteration. Thus,  no information exchange is needed.

\item
Computing the second term
$\mathcal{I}_{m,k}^{(n),\texttt{out},j}=\sum_{u \in \mathcal{B}_{j}^{(n)}}\left|  \vec{\mathbf{h}}_{j,k}^{(n)}\mathbf{w}_{j,u}^{(n)}\right| ^{2},  j \in \mathcal{M}\backslash m$ in (\ref{Interference}). BS $j \in \mathcal{M}\backslash m$ needs the local beam-vectors $\{\mathbf{w}_{j,u}^{(n)}, u \in \mathcal{B}_{j}^{(n)}\}$ computed at the previous iteration, and the interference downlink channel $\mathbf{h}_{j,k}^{(n)}$, which is sent from user $k \in \mathcal{B}_{m}^{(n)}$ to the serving BS $m$, and then from BS $m$ to BS $j$.
BS $j$ calculates  $\mathcal{I}_{m,k}^{(n),\texttt{out},j}$ and then sends it to BS $m$.
\end{enumerate}

BS $m$ can compute $\mathbf{L}_{m,k}^{(n)}$ in (\ref{LeakMatrix}) for each user $k \in \mathcal{B}_{m}^{(n)}$ and each sub-channel $n \in \mathcal{N}$ through the following.

\begin{enumerate}
\item
Computing the first term $\mathbf{L}_{m,k}^{(n),\texttt{in}}=\sum_{\substack{k' \in \mathcal{B}_m^{(n)}
 \backslash k}}\pi_{m,k'}^{(n)} \mathbf{h}_{m,k'}^{(n)}\vec{\mathbf{h}}_{m,k'}^{(n)}$ in (\ref{LeakMatrix}). BS $m$ only needs the direct downlink channels $\{\mathbf{h}_{m,k'}^{(n)}, k' \in \mathcal{B}_{m}^{(n)}\backslash k\}$ and
the local interference pricing rates $\{\pi_{m,k'}^{(n)}, k' \in \mathcal{B}_{m}^{(n)}\backslash k\}$ computed at the previous
iteration. Thus no information exchange is needed.

\item
Computing the second term
$\mathbf{L}_{m}^{(n),\texttt{out},j}=\sum_{u \in \mathcal{B}_{j}^{(n)}}\pi_{j,u}^{(n)} \mathbf{h}_{m,u}^{(n)}\vec{\mathbf{h}}_{m,u}^{(n)},  j \in \mathcal{M}\backslash m$  in (\ref{LeakMatrix}). BS $j$ needs the interference downlink channels $\{\mathbf{h}_{m,u}^{(n)}, u \in \mathcal{B}_{j}^{(n)}\}$, each of which is sent from user $u$ to the serving BS $j$. Notice that  these channels   have already been sent when computing $\mathcal{I}_{j,u}^{(n),\texttt{out},m}$. Thus this incurs no additional information exchange. On the other hand, the interference pricing rates  $\{\pi_{j,u}^{(n)}, u \in \mathcal{B}_{j}^{(n)}\}$ can be computed  locally according to (\ref{Equ_price}) for which the quantity $\mathcal{I}_{j,u}^{(n)}$ is needed, which in turn has been computed at the previous iteration.
After computing  $\{\pi_{j,u}^{(n)}, u \in \mathcal{B}_{j}^{(n)}\}$, BS $j$ sends them to BS $m$.
\end{enumerate}

{\bf Remark}: Note that compared with \cite{venturino:2010}, in our scheme,  BS $j$ sends $\{\pi_{j,u}^{(n)}, u \in \mathcal{B}_{j}^{(n)}\}$ instead of  $\mathbf{L}_{m}^{(n),\texttt{out},j}$ itself to BS $m$. Since $\mathbf{L}_{m}^{(n),\texttt{out},j}$ is a $T \times T$ complex-valued matrix, while $\{\pi_{j,u}^{(n)}, u \in \mathcal{B}_{j}^{(n)}\}$ is a $Q \times 1$ real-valued vector, and
typically $Q \leq T$, our scheme incurs a much lower information exchange overhead.


\section{Per-BS Beam-vector Update\\ via Dual Decomposition}
\label{dd.sec}

In this section, we derive a dual decomposition algorithm for obtaining the solution to the KKT conditions of Problem  (\ref{BSSolver}) at each BS, which is the key step in
Algorithm \ref{DPSPB} (line 5 and 6).
\subsection{Dual Decomposition}
The dual decomposition technique is an effective method for
decoupling the coupled constraints and performing distributed
optimization.

First, by introducing a set of scalar variables
$\mathbf{p}_{m}=\left\{ {p_{m,k}^{(n)}, k \in \mathcal{B}_m^{(n)}, n
\in \mathcal{N}} \right\}$, we can rewrite the optimization problem
(\ref{BSSolver}) as follows:
\begin{equation}\label{BSSolver2}
\begin{split}
 \max_{ \mathbf{W}_m } & \ \ \sum_{n \in \mathcal{N}} \sum_{k \in
\mathcal{B}_m^{(n)}} \Big (U_{m,k}^{(n)}
(\Gamma_{m,k}^{(n)})- \vec{\mathbf{w}}_{m,k}^{(n)}\mathbf{L}_{m,k}^{(n)}\mathbf{w}_{m,k}^{(n)} \Big ),\\
 \mathrm{s.t.} & \ \  \sum_{n \in \mathcal{N}}\sum_{k \in \mathcal{B}_m^{(n)}} p_{m,k}^{(n)}\leq  P_m, \\
&  \ \  \vec{\mathbf{w}}_{m,k}^{(n)}\mathbf{w}_{m,k}^{(n)}  \le
p_{m,k}^{(n)},\ \; k \in \mathcal{B}_m^{(n)}, \; n \in \mathcal{N}.
\end{split}
\end{equation}
Notice that the optimization problem (\ref{BSSolver2}) has only one
single coupled constraint $\sum_{n \in \mathcal{N}}\sum_{k \in
\mathcal{B}_m^{(n)}} p_{m,k}^{(n)}\leq   P_m$. Then, we form the
Lagrangian (\ref{Lagrangian1}) of the optimization problem (\ref{BSSolver2}) with
respect to the coupled constraint. In (\ref{Lagrangian1}), $\lambda _m$ denotes the Lagrangian dual variable.
\begin{figure*}[!t]
\begin{equation}\label{Lagrangian1}
\begin{split}
\;\;\;\;\;\  \tilde L_m (\mathbf{W}_{m}, \mathbf{p}_{m},\lambda _m )
&= \sum_{n \in \mathcal{N}} \sum_{k \in \mathcal{B}_m^{(n)}} \Big (U_{m,k}^{(n)}(\Gamma_{m,k}^{(n)})  - \vec {\mathbf{w}}_{m,k}^{(n)} \mathbf{L}_{m,k}^{(n)} \mathbf{w}_{m,k}^{(n)} \Big)
  - \lambda _m \Big( {\sum_{n \in \mathcal{N}} {\sum_{k \in \mathcal{B}_m^{(n)}} {p_{m,k}^{(n)} } }  - P_m } \Big) \\
&= \sum_{n \in \mathcal{N}} \sum_{k \in \mathcal{B}_m^{(n)}} \Big (U_{m,k}^{(n)}(\Gamma_{m,k}^{(n)})  - \vec {\mathbf{w}}_{m,k}^{(n)} \mathbf{L}_{m,k}^{(n)} \mathbf{w}_{m,k}^{(n)}  - \lambda _m p_{m,k}^{(n)}  \Big)  + \lambda _m  P_m,
 \end{split}
\end{equation}
\hrulefill
\end{figure*}

Define the dual problem as
\begin{equation}\label{BSmaster2}
\mathop {\min }\limits_{\lambda _m } D_m \left( {\lambda _m }
\right),
\end{equation}
where the objective function $D_m \left( {\lambda _m } \right)$  is
\begin{equation}\label{Dualproblem}
\begin{split}
  \max_{\mathbf{W}_{m}, \mathbf{p}_{m}}   & \ \ \tilde L_m (\mathbf{W}_{m}, \mathbf{p}_{m},\lambda _m )
 \\
  \mathrm{s.t.} & \ \   \vec{\mathbf{w}}_{m,k}^{(n)}\mathbf{w}_{m,k}^{(n)}  \le p_{m,k}^{(n)}, \ k \in \mathcal{B}_m^{(n)}, \ n \in
  \mathcal{N}.
\end{split}
\end{equation}
Notice that the dual function $D_m \left( {\lambda _m } \right)$ is the pointwise maximum of
a family of affine functions of  $\lambda _m$, hence it is a convex function of $\lambda _m$.

\subsection{Decoupled Subproblems}

First, we need to compute  $D_m \left( {\lambda _m } \right)$ for a
fixed $\lambda _m$. Due to its separable
structure, the dual function  $D_{m}$ can be decomposed into $NQ$
subproblems $D_{m,k}^{(n)}, n \in \mathcal{N}, k \in
\mathcal{B}_m^{(n)}$ as follows:
\begin{eqnarray}\label{BSsubproblem}
&& \max_{\left\{ {\mathbf{w}_{m,k}^{(n)}
,p_{m,k}^{(n)} } \right\}} \ \ U_{m,k}^{(n)}(\Gamma_{m,k}^{(n)})  -
\vec {\mathbf{w}}_{m,k}^{(n)} \mathbf{L}_{m,k}^{(n)}
\mathbf{w}_{m,k}^{(n)}  - \lambda _m p_{m,k}^{(n)}, \nonumber \\
 & & \;\;\;\;\;\;\; \mathrm{s.t.} \ \ \ \;\;\;\;\;\;  \vec {\mathbf{w}}_{m,k}^{(n)} \mathbf{w}_{m,k}^{(n)}  \le
 p_{m,k}^{(n)}.
\end{eqnarray}
The Lagrangian of the subproblem $D_{m,k}^{(n)}$ is given by
\begin{eqnarray}
&\hspace{-2.5cm}\widehat{L}_{m,k}^{(n)} \left( {\mathbf{w}_{m,k}^{(n)} ,p_{m,k}^{(n)} ,\lambda _m ,\nu _{m,k}^{(n)} } \right)   \nonumber\\
&\hspace{-1cm}=U_{m,k}^{(n)} \left( {\Gamma _{m,k}^{(n)} } \right) - \vec
{\mathbf{w}}_{m,k}^{(n)} \mathbf{L}_{m,k}^{(n)}
\mathbf{w}_{m,k}^{(n)}  - \lambda _m p_{m,k}^{(n)} \\
 &\hspace{-2.5cm}-\nu_{m,k}^{(n)} \left( {\vec {\mathbf{w}}_{m,k}^{(n)}
\mathbf{w}_{m,k}^{(n)}  - p_{m,k}^{(n)} } \right),\nonumber
\end{eqnarray}
where $\nu _{m,k}^{(n)}$ is a dual variable associated with the
constraint $\vec{\mathbf{w}}_{m,k}^{(n)} \mathbf{w}_{m,k}^{(n)} \leq p_{m,k}^{(n)}$.\\

We next obtain the KKT conditions, given by
\begin{eqnarray}
  \frac{{\partial \widehat{L}_{m,k}^{(n)} }}{{\partial \mathbf{w}_{m,k}^{(n)} }} &=&    {U_{m,k}^{(n)} \Big( \Gamma_{m,k}^{(n)} \Big)} ^\prime \cdot \frac{{2\vec{\mathbf{h}}_{m,k}^{(n)}\mathbf{h}_{m,k}^{(n)} \mathbf{w}_{m,k}^{(n)} }}{{1 + \mathcal{I}_{m,k}^{(n)} }} \nonumber\\ &&- 2\mathbf{L}_{m,k}^{(n)} \mathbf{w}_{m,k}^{(n)}  - 2\nu _{m,k}^{(n)} \mathbf{I}_T  \mathbf{w}_{m,k}^{(n)}  = 0  \\
  \frac{{\partial \widehat{L}_{m,k}^{(n)} }}{{\partial p_{m,k}^{(n)} }} &=&  - \lambda _m  + \nu _{m,k}^{(n)}  = 0
\end{eqnarray}
where $\mathbf{I}_T$ indicates the $T \times T$ identity matrix.

By combining the above two equations, we obtain
\begin{eqnarray}\label{KKT}
   &\hspace{-0.5cm}{U_{m,k}^{(n)} \Big( {\frac{{\big| {\vec {\mathbf{h}}_{m,k}^{(n)} \mathbf{w}_{m,k}^{(n)} } \big|^2 }}{{1 + \mathcal{I}_{m,k}^{(n)} }}}
 \Big)}  ^\prime  \frac{{\mathbf{h}_{m,k}^{(n)} \vec {\mathbf{h}}_{m,k}^{(n)} \mathbf{w}_{m,k}^{(n)} }}{{1 + \mathcal{I}_{m,k}^{(n)} }}
  =\mathbf{T}_{m,k}^{(n)} \mathbf{w}_{m,k}^{(n)}, \\
&\hspace{-4.3cm}{\rm with} \ \ \ \mathbf{T}_{m,k}^{(n)} = \mathbf{L}_{m,k}^{(n)}
+ \lambda _m \mathbf{I}_T.
\end{eqnarray}
Solving $\mathbf{w}_{m,k}^{(n)   }$ from  (\ref{KKT}), we obtain the expression for the
beam-vectors associated with user $k \in \mathcal{B}_{m}^{(n)}$ for a fixed $\lambda_m$, as follows.
\begin{proposition}
For a fixed
$\lambda _m \geq 0$, the solution to the KKT conditions of problem (\ref{BSsubproblem}) is of the following form\footnote{$(\cdot)^{\dag}$ denotes the pseudo-inverse; ${\rm Inv}\{U'\}$ is the inverse function of $U'$.}:
\begin{eqnarray}\label{KKTsolution}
 &\hspace{-1cm}\mathbf{w}_{m,k}^{(n) * } = \mathbf{T}_{m,k}^{(n) \dag }  \mathbf{h}_{m,k}^{(n)} \sqrt {\left( {1 + \mathcal{I}_{m,k}^{(n)}
 } \right)\Phi _{m,k}^{(n)} \Upsilon _{m,k}^{(n)} }, \\
 \label{KKTsolution2}
&\hspace{-0.8cm}p_{m,k}^{(n) * } =\vec {\mathbf{w}}_{m,k}^{(n) * }\mathbf{w}_{m,k}^{(n) * }  = \left( {1 + \mathcal{I}_{m,k}^{(n)} }
\right)\Phi _{m,k}^{(n)} \Psi _{m,k}^{(n)}, \\
&\hspace{-1cm}{\rm with} \ \
\Phi _{m,k}^{(n)} = {\rm Inv} {\Big\{   U_{m,k}^{(n)} \Big( {\frac{{1 + \mathcal{I}_{m,k}^{(n)} }}{{\vec {\mathbf{h}}_{m,k}^{(n)} \mathbf{T}_{m,k}^{(n) \dag }  \mathbf{h}_{m,k}^{(n)} }}} \Big)^\prime}  \Big\}  ,\\
&\hspace{-1.1cm}\Upsilon _{m,k}^{(n)}  = 1/{\left( {\vec {\mathbf{h}}_{m,k}^{(n)} \mathbf{T}_{m,k}^{(n) \dag }  \mathbf{h}_{m,k}^{(n)} } \right)^2 },\\
&\hspace{-1.55cm}\Psi _{m,k}^{(n)}  = \Big\| {\mathbf{T}_{m,k}^{(n) \dag }  \mathbf{h}_{m,k}^{(n)} } \Big\|^2 \Upsilon _{m,k}^{(n)}.
\end{eqnarray}
Furthermore,  if $\mathbf{w}_{m,k}^{(n) * }\neq {\bf 0}$, $\lambda _m=0$ is feasible only if $\mathbf{h}_{m,k}^{(n)} \in \varrho(\mathbf{L}_{m,k}^{(n)})$, where $\varrho(\mathbf{L}_{m,k}^{(n)})$ denotes the column span of the matrix
$\mathbf{L}_{m,k}^{(n)}$.
 \end{proposition}
{\em Proof:} \ See   Appendix B. \hfill $\Box$



\subsection{Master Problem}

We need to solve the master problem (\ref{BSmaster2}) on
top of the $NQ$ subproblems. Since the master problem is   convex in $\lambda_m$,
we will apply the subgradient method. Notice that
\begin{figure*}[!t]
\begin{eqnarray}
 D_m \left( {\tilde\lambda_m } \right)
  &=& \mathop {\max }\limits_{\mathbf{W}_{m}, \mathbf{p}_{m}} \sum_{n \in \mathcal{N}}\sum_{k \in \mathcal{B}_m^{(n)}} {\left\{ {U_{m,k}^{(n)}(\Gamma_{m,k})  - \vec {\mathbf{w}}_{m,k}^{(n)}\mathbf{L}_{m,k}^{(n)}\mathbf{w}_{m,k}^{(n)} } \right\}}  -
  \tilde\lambda_m \Big( {\sum_{n \in \mathcal{N}} {\sum_{k \in \mathcal{B}_m^{(n)}} {p_{m,k}^{(n)} } }  -  P_m } \Big)
  \nonumber \\
  &\ge& \sum_{n \in \mathcal{N}}\sum_{k \in \mathcal{B}_m^{(n)}} {\left\{ {U_{m,k}^{(n)}  - \vec {\mathbf{w}}_{m,k}^{(n)*} \mathbf{L}_{m,k}^{(n)}\mathbf{w}_{m,k}^{(n)*} } \right\}}   - \tilde\lambda_m \Big( {\sum_{n \in \mathcal{N}} {\sum_{k \in \mathcal{B}_m^{(n)}} {p_{m,k}^{(n)*} } }  -  P_m } \Big) \nonumber\\
  &=& D_m \left( {\lambda _m } \right) - \Big( {\sum_{n \in \mathcal{N}}\sum_{k \in \mathcal{B}_m^{(n)}} {p_{m,k}^{(n)*} }   -  P_m } \Big)\left( {\tilde\lambda_m  - \lambda _m } \right),
\end{eqnarray}
\hrulefill
\end{figure*}
where $ \mathbf{W}_{m}^* = \{\mathbf{w}_{m,k}^{(n)*}, n \in \mathcal{N}, k \in \mathcal{B}_m^{(n)} \}$ is
 the optimizer for (\ref{BSmaster2}) in the definition of
$D_m \left( {\lambda _m } \right)$.

Thus, we can set the subgradient of $D_m \left( {\lambda _m } \right)$ as
$ P_m  - \sum_{n \in \mathcal{N}} {\sum_{k \in \mathcal{B}_m^{(n)}} {p_{m,k}^{(n)*} } }$.
The subgradient search suggests that we should increase $\lambda _m  $
if $P_m  < \sum_{n \in \mathcal{N}} {\sum_{k \in \mathcal{B}_m^{(n)}} {p_{m,k}^{(n)*} } }$; and decrease $\lambda _m $
otherwise.
Notice that the adjustment
occurs in a one-dimensional space, thus a simple bisection method can be employed.

\subsection{The Beamforming Algorithm at Each BS}
Finally  we  summarize in Algorithm 2 the dual-decomposition-based
beamforming algorithm at each BS.


\begin{algorithm}
Initialize  $\lambda _m^{\min }$ and $\lambda _m^{\max }$;\\
\textsf{Repeat}\\
$\;\;\;\; $ Set ${\lambda _m  \leftarrow \left( {\lambda _m^{\min}  + \lambda _m^{\max } } \right)/2}$;\\
$\;\;\;\; $ \textsf{Repeat} through $k = 1,...,Q;1,...,Q;...$\\
$\;\;\;\; \;\; $ Compute $\{\mathcal{I}_{m,k}^{(n),\mathrm{in}}\}_{n=1}^{N}$ and thus $\{\mathcal{I}_{m,k}^{(n)}\}_{n=1}^{N}$\\
$\;\;\;\; \;\; $ according to (\ref{Interference});\\
$\;\;\;\; \;\; $ Compute $\{\mathbf{L}_{m,k}^{(n),\mathrm{in}}\}_{n=1}^{N}$ and thus $\{\mathbf{L}_{m,k}^{(n)}\}_{n=1}^{N}$ \\
$\;\;\;\; \;\; $ according to (\ref{LeakMatrix});\\
$\;\;\;\; \;\; $ Compute $\{\mathbf{w}_{m,k}^{(n),*}, p_{m,k}^{(n),*}\}_{n=1}^{N}$ for (\ref{BSsubproblem})  according to\\
$\;\;\;\; \;\; $ (\ref{KKTsolution}) and (\ref{KKTsolution2}).\\
$\;\;\;\; $ \textsf{Until} convergence\\
$\;\;\;\; $ \textsf{If} $\sum_{n \in \mathcal{N}} {\sum_{k \in \mathcal{B}_m^{(n)}} {p_{m,k}^{(n)*} } } > P_m $\\
$\;\;\;\; \;\; $ \textsf{Then} set  $\lambda _m^{\min } \leftarrow {\lambda _m}$\\
$\;\;\;\; \;\; \;$ \textsf{Else} set  $\lambda _m^{\max } \leftarrow {\lambda _m}$\\
$\;\;\;\; $ \textsf{End If}\\
\textsf{Until} $|\lambda _m^{\max }-\lambda _m^{\min }|\rightarrow 0$\\
\caption{An algorithm for updating the beam-vectors at each BS $m$. }
\label{PSIB}
\end{algorithm}


Note that  Algorithm \ref{PSIB} can also be viewed as an iterative procedure
for solving the KKT conditions of  problem (\ref{BSSolver}),
which consists of the following.
\begin{enumerate}
\item The stationarity condition (\ref{KKT})
for $k \in \mathcal{B}_{m}^{(n)}$, $n \in \mathcal{N}$.

\item The sum-power constraint
\begin{equation}\label{powerconstraint}
\sum_{n \in \mathcal{N}}\sum_{k \in \mathcal{B}_m^{(n)}} \vec{\mathbf{w}}_{m,k}^{(n)}\mathbf{w}_{m,k}^{(n)}\leq   P_m.
\end{equation}

\item The complementary slackness
conditions:
\begin{equation}\label{slackness}
\lambda_{m}(\sum_{n \in \mathcal{N}}\sum_{k \in \mathcal{B}_m^{(n)}} \vec{\mathbf{w}}_{m,k}^{(n)}\mathbf{w}_{m,k}^{(n)}- P_m)=0,
\end{equation}
with $\lambda_{m}\geq 0$.

\end{enumerate}
Starts with a given $\lambda_{m}$, Algorithm \ref{PSIB} solves
(\ref{KKT}) for $k \in \mathcal{B}_{m}^{(n)}$, $n \in \mathcal{N}$, and then adjusts $\lambda_{m}$ according to the search direction suggested by the
power constraint (\ref{powerconstraint}), such as to satisfy the complementary slackness
conditions (\ref{slackness}).

Notice that since $\mathcal{N} = \{1, 2, . . . ,N\}$ is a set of orthogonal sub-channels,
for $n_{1}, n_{2} \in \mathcal{N}, n_{1}\neq n_{2}$, and $k_{1}, k_{2} \in \mathcal{B}_m^{(n)}, k_{1}\neq k_{2}$,
we can update $\{\mathbf{w}_{m,k_{1}}^{(n_{1})},p_{m,k_{1}}^{(n_{1})}\}$
for the subproblem  $D_{m,k_{1}}^{(n_{1})}$,
and $\{\mathbf{w}_{m,k_{2}}^{(n_{2})},p_{m,k_{2}}^{(n_{2})}\}$
for the subproblem  $D_{m,k_{2}}^{(n_{2})}$ in parallel.
Such a simultaneous update can improve the convergence speed of the iterative procedure, especially  when $N$ is  large.

Moreover,
it is easy to derive the following KKT
conditions of the original problem (\ref{Equ4}):
\begin{eqnarray}\label{KKT2}
   &\hspace{-5cm}U_{m,k}^{(n)} \Big( {\frac{{\big| {\vec {\mathbf{h}}_{m,k}^{(n)} \mathbf{w}_{m,k}^{(n)} } \big|^2 }}{{1 + \mathcal{I}_{m,k}^{(n)} }}}
 \Big)^\prime  \frac{{\mathbf{h}_{m,k}^{(n)} \vec {\mathbf{h}}_{m,k}^{(n)} \mathbf{w}_{m,k}^{(n)} }}{{1 + \mathcal{I}_{m,k}^{(n)} }}
  = \mathbf{T}_{m,k}^{(n)} \mathbf{w}_{m,k}^{(n)}, \nonumber
  &\hspace{-1.5cm}\\ k \in \mathcal{B}_{m}^{(n)}, n \in \mathcal{N}, m \in \mathcal{M}, \\
 \label{powerconstraint2}
&\hspace{-5.7cm} \sum_{n \in \mathcal{N}}\sum_{k \in \mathcal{B}_m^{(n)}} \vec{\mathbf{w}}_{m,k}^{(n)}\mathbf{w}_{m,k}^{(n)} \leq   P_m, \ \;m \in \mathcal{M},\\
 \label{slackness2}
&\hspace{-4.5cm}\lambda_{m}\Big(\sum_{n \in \mathcal{N}}\sum_{k \in \mathcal{B}_m^{(n)}} \vec{\mathbf{w}}_{m,k}^{(n)}\mathbf{w}_{m,k}^{(n)}- P_m
\Big)=0, \ \; \hspace{-0.2cm}m \in \mathcal{M}.
\end{eqnarray}
Let
$\mathbf{W}_m^{KKT}$ the beam-vector set satisfying
the KKT conditions of (\ref{BSSolver}) for   BS $m \in \mathcal{M}$. By comparing (\ref{KKT}), (\ref{powerconstraint}), (\ref{slackness}) with (\ref{KKT2}), (\ref{powerconstraint2}), (\ref{slackness2}), respectively, it is obvious that
$\mathbf{W}^{KKT}=\{\mathbf{W}_1^{KKT},\ldots,\mathbf{W}_{M}^{KKT}\}$ is the beam-vector set satisfying the KKT
conditions of the original problem (\ref{Equ4}).

In Algorithm 2, each BS $m \in \mathcal{M}$ can achieve the KKT solution to (\ref{BSSolver}).
From Proposition 1, we know that Algorithm 1   converges to an NE point.
Furthermore, line 6 in Algorithm 1 guarantees that the beam-vector  update  at each BS cannot decrease the total utility.
Thus, the total utility at the NE point is {\it not} smaller than that at the point satisfying the KKT
conditions of the original problem (\ref{Equ4}).

\section {Simulation Results}

\subsection {Simulation Setup}
 \begin{figure}[htbp]
 \centering
  \includegraphics[scale=0.45,bb=108 150 434 495]{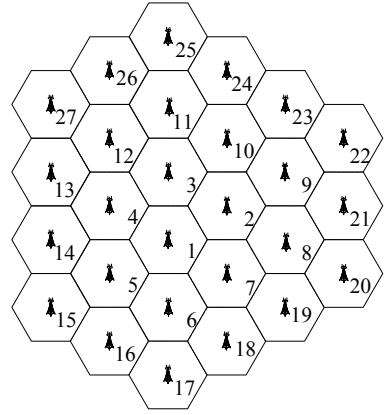}
    \centering
    \caption{The simulated network.}
 \label{CellConfigure}
 \end{figure}
We consider a network with hexagonal cells $\mathcal{M}_{t}=\{1,...,27\}$ shown in Fig.~\ref{CellConfigure}.
The distance between adjacent
BSs is $D_{BS}=$2000m. Let $\mathcal{M}_{co}=\{1,...,M\}$ be the set of
coordinated BSs, and $\mathcal{M}_{un}=\mathcal{M}_{t} \backslash\mathcal{M}_{co}$
be the set of uncoordinated BSs. On each sub-channel, $Q$ users are uniformly displaced around the
serving BS within a circular annulus of external and internal radii of $D$ and $0.9D$, respectively.
Since the proposed method is expected to benefit most the cell-edge users,
by setting $D=$1000m as the default value all users are around the cell edges.

The base-band fading channel from the $m$-th
BS to the $k$-th user on sub-channel $n$ is modeled as
\begin{equation}\label{ChanelModel}
{\mathbf{h}}_{m,k}^{(n)}=\left(\frac{200}{d_{m,k}^{(n)}}\right) ^{3.5}l_{m,k}^{(n)}\bar{\mathbf{h}}_{m,k}^{(n)},
\end{equation}
where $d_{m,k}^{(n)}$ is the distance  from the $m$-th
BS to the $k$-th user on sub-channel $n$; $10\log_{10}l_{m,k}^{(n)}$ is
a real Gaussian random variable with zero mean and a standard deviation of 8 accounting for
the large scale
log-normal shadowing; finally, $\bar{\mathbf{h}}_{m,k}^{(n)}\sim {\mathcal N}_c (\mathbf{0}_{T},\mathbf{I}_{T})$ is a circularly symmetric complex
Gaussian random vector accounting for Rayleigh fast fading.

The total noise power $\mathcal{\eta}_{m,k}^{(n)}$ at each user is modeled as
\begin{equation}\label{NoiseModel}
\mathcal{\eta}_{m,k}^{(n)}=\sigma^{2}+\sum_{m \in \mathcal{M}_{un}} \left(\frac{200}{d_{m,k}^{(n)}}\right) ^{3.5}l_{m,k}^{(n)}\frac{ P}{N},
\end{equation}
where  $\sigma^{2}$ is the thermal noise power, and the second term
accounts for the uncoordinated inter-cell interference. Note that
$\mathcal{\eta}_{m,k}^{(n)}$ is used to obtain the normalized signal
model (\ref{Equ1}). We assume equal power  (i.e., $P_{m}=P$) for
each BS $m$ in the following simulation, and consider the system the
performance under different signal-to-noise ratio (SNR), which is
defined as $\gamma \triangleq {P}/\sigma^{2}$.
\subsection {Convergence Behavior}
\label{sec:Convergence}
We first illustrate the convergence behavior of the proposed distributed beamforming scheme,  under three different utility functions, namely,
\begin{enumerate}
\item Proportional fairness utility: $U_{m,k}^{(n)}(\Gamma_{m,k}^{(n)}) = \frac{1}{NM}\log_2(\Gamma_{m,k}^{(n)});$
\item $\alpha$-fairness utility: $U_{m,k}^{(n)}(\Gamma_{m,k}^{(n)}) = \frac{1}{NM}\frac{(\Gamma_{m,k}^{(n)})^{(1-\alpha)}}{1-\alpha}$ with $(\alpha=2);$
\item Sum-rate utility: $U_{m,k}^{(n)}(\Gamma_{m,k}^{(n)}) =  \frac{1}{NM}\log_2(1+\Gamma_{m,k}^{(n)}).$
\end{enumerate}
The number of coordinated cells is $M=7$; the number of sub-channels is $N=3$; the number of
transmit antennas at each BS is $T=6$; the number of  SDMA users is $Q=3$; and the location
parameter of the users is $D=1000$m. Algorithm 1 is initialized by the channel-matched (CM)
beamformers.
Note that initializing with the more sophisticated beamformers, such as the
in-cell zero-forcing (ICZF) beamformer, may only slightly increase the convergence speed.

 \begin{figure}[htbp]
\centering
\includegraphics[scale=0.65,bb=87 464 476 769]{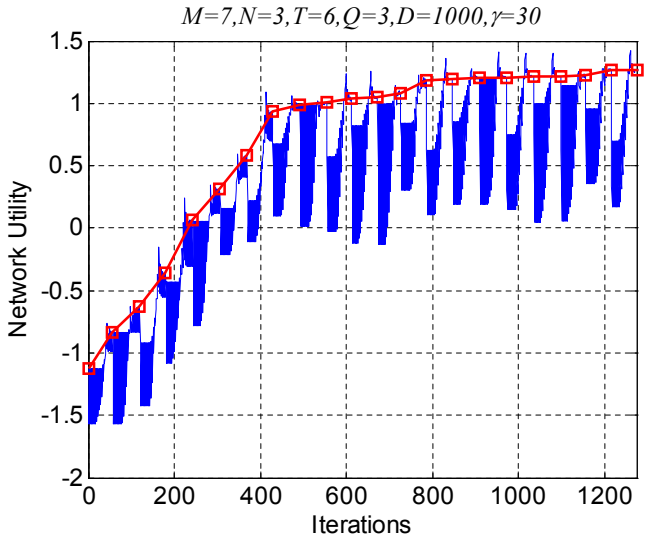}
\centering
\caption{Convergence of Algorithms 1 \& 2 (proportion fairness utility).}
\centering
\label{proportion_fair_convergence}
\centering
\end{figure}

\begin{figure}[htbp]
\centering
\includegraphics[scale=0.65,bb=87 464 466 769]{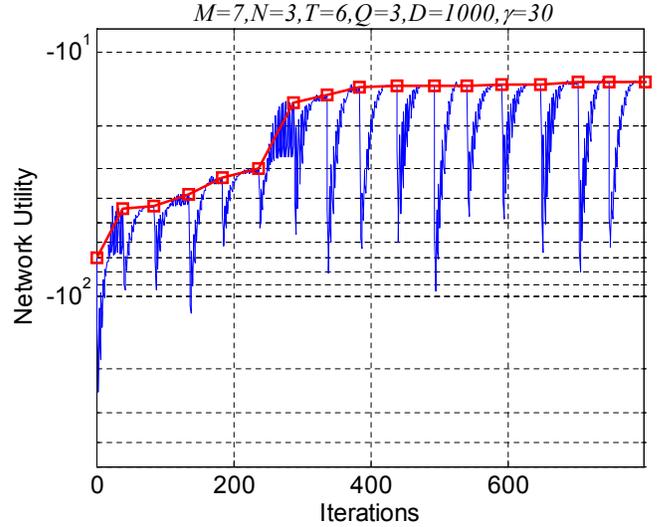}
\centering
\caption{Convergence of Algorithms 1 \& 2 ($\alpha$-fairness utility).}
\centering
\label{alpha_fair_convergence}
\centering
\end{figure}

\begin{figure}[htbp]
\centering
\includegraphics[scale=0.65,bb=87 464 476 769]{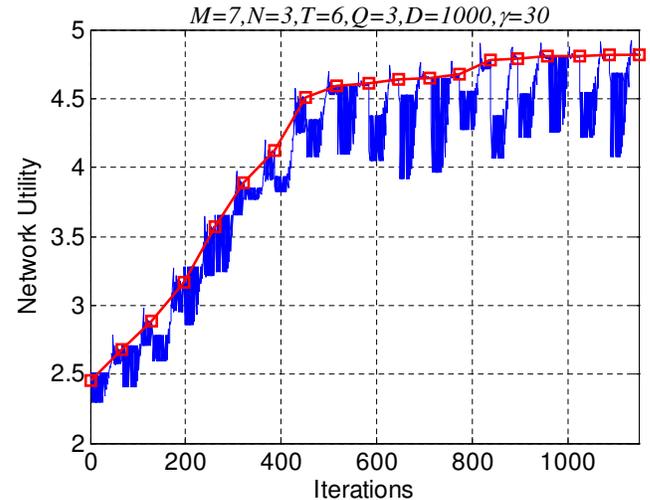}
\centering
\caption{Convergence of Algorithms 1 \& 2 (weighted sum-rate utility).}
\centering
\label{weight_sum_rate_convergence}
\centering
\end{figure}

The convergence behaviors of Algorithm 1 and Algorithm 2  at SNR $\gamma=30\mathrm{dB}$ are shown in
Figs.~\ref{proportion_fair_convergence}-\ref{weight_sum_rate_convergence}
for the above three utility functions, respectively.
 In these figures, the squares correspond to the outer iteration, i.e., the iterations of Algorithm 1.  In each outer iteration,   one BS
  updates its  beam-vectors   for  all its users. The solid lines correspond to the inner iterations, i.e., the iterations of Algorithm 2. In each inner iteration, the BS
 updates the beam-vectors of one of its users.
It is seen that the proposed distributed beamforming technique can
significantly improve  upon the initial network utility   through optimizing the   power
allocation across the beams and   the beam directions according
to the conditions of the in-cell and out-cell interference.
Moreover, the value of the network utility monotonically increases at each outer iteration,
which confirms the theoretic result of Proposition 1.

\begin{figure}
\centering
\includegraphics[scale=0.65,bb=87 464 476 769]{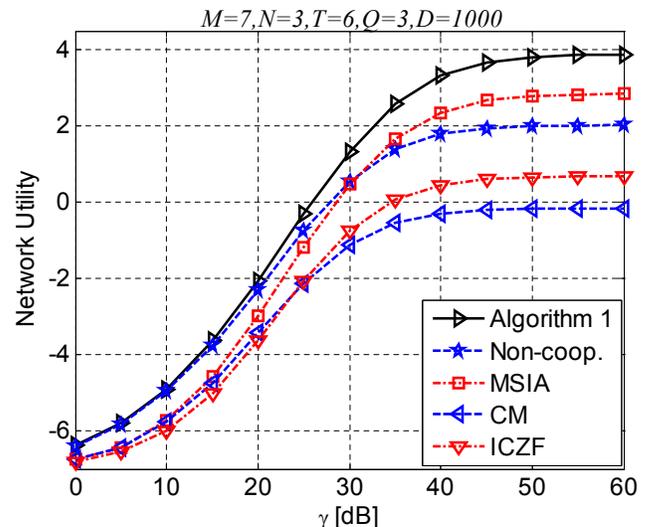}
\centering
\caption{Total proportional fairness  utility   versus SNR.}
\centering
 \label{proportion_fair_gain}
\centering
\end{figure}

\begin{figure}
\centering
\includegraphics[scale=0.65,bb=87 464 476 769]{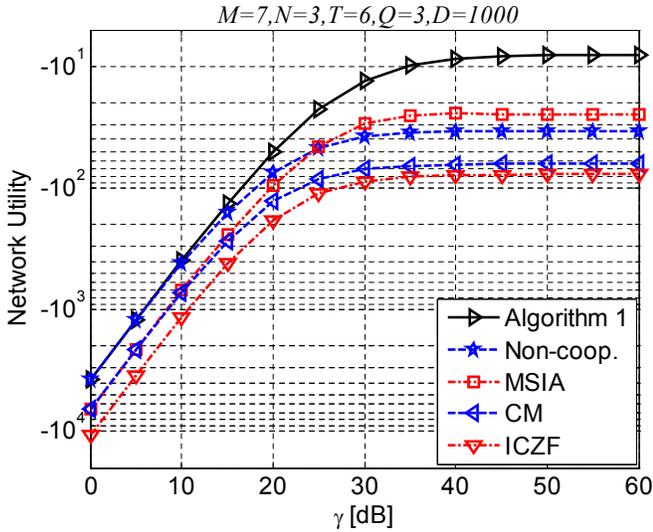}
\centering
\caption{Total $\alpha$-fairness utility versus SNR.}
\centering
\label{alpha_fair_gain}
\centering
\end{figure}

\begin{figure}
\centering
\includegraphics[scale=0.65,bb=87 464 476 769]{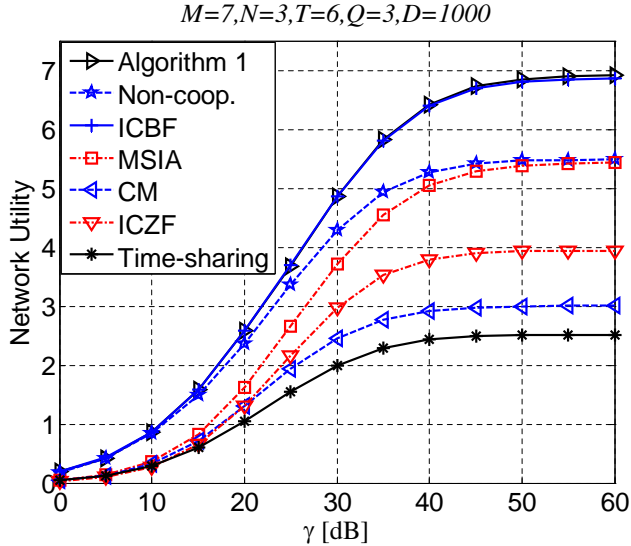}
\centering
\caption{Total weighted sum-rate utility versus SNR.  }
\centering
\label{weight_sum_rate_gain}
\centering
\end{figure}

Our extensive simulations reveal  that the convergence speed of Algorithm 1 is
affected by the system parameters as follows.
\begin{itemize}
\item  The number of coordinated BSs $M$:
A larger $M$ corresponds to a slower convergence.

\item  The SNR $\gamma$: A larger $\gamma$ corresponds to a slower convergence.

\item The number of antennas $T$ and the number of co-channel users $Q$:
They only slightly affect the convergence of the outer iteration, i.e., Algorithm 1; but significantly affect the convergence of the
 inner iteration, i.e., Algorithm 2. Specifically, larger $T$ and $Q$ correspond to slower convergence of Algorithm 2.

\end{itemize}

In general it is seen that a relatively small number of
outer iterations  are sufficient for Algorithm 1 to  converge.
Note that there is no   step-size parameter in Algorithm 1 and
 the overall utility could change
dramatically in a single update, leading to rapid convergence.
This is in contrast to some conventional
  distributed optimization
algorithms, in which some step-size parameter controls
the speed of convergence.

\subsection {Performance Comparisons}
We next compare the performance of the proposed distributed beamforming method with some
existing techniques, including the simple channel matching (CM) method, the in-cell zero-forcing
(ICZF) method, the iterative coordinated beam-forming (ICBF) method proposed in \cite{venturino:2010}, and the more recent maximum SINR
interference alignment (MSIA) method \cite{Gomadam2009}, as well as the approach based on the pure non-cooperative
game (i.e., without the pricing mechanism) and  the full-cooperation based method. Furthermore, the sum-rate performance of the time-sharing
scheme  is also considered,  where the $Q$ users in each cell access each
sub-channel via TDMA.

In Figs.~\ref{proportion_fair_gain}-\ref{weight_sum_rate_gain}, the total utility values versus the SNR for the above-mentioned methods are plotted
for the three utility functions, respectively. Note that for CM, ICZF and MSIA, the beamformer
solutions are independent of the utility function. It is seen that the proposed distributed
beamforming method outperform all other techniques in the sense of offering higher total network
utilities.
Moreover, we note that the network utility gain provided by Algorithm 1 is affected by the
system parameters as follows.
\begin{itemize}

\item The distance from the  BS to the user $D$:
The network utility gain provided by Algorithm 1
is  larger when the users are closer to the cell edge.
Intuitively, the cell-edge users experience higher path losses and suffer from  higher out-cell
interference. Through Algorithm 1, both the available power
across beams  and the beam directions are optimized to mitigate these effects.
On the other hand, when the users are  close to the serving BS and away
from the cell-edge, the per-cell optimized solution based on the
non-cooperative game can achieve high performance without any information
exchange among the BSs.

\item The number of antennas $T$ and the number of cochannel users $Q$:
 The network utility gain
provided by Algorithm 1 is larger for larger $T$ and $Q$.
Intuitively, when the number  of co-channel users $Q$  is large, the users suffer from the high out-cell and in-cell
interference.
Due to the large number of antennas,  Algorithm 1 has
enough degrees of freedom  to ensure the good quality of service
of each user while causing minimum amount of in-cell and out-cell interference.

\item The number of coordinated BSs $M$:
The performance gain provided by Algorithm 1 becomes larger
for larger $M$. This is because a larger $M$ corresponds to a larger
number of degrees of freedom to mitigate the interference.

\item The signal-to-noise ratio $\gamma$:
Algorithm 1 only provides marginal gains at low SNR, while the gain is more
prominent at high SNR.
This is because at high SNR, the interference becomes dominant factor for
limiting the system performance, which can be effectively mitigated by
Algorithm 1.
\end{itemize}


\medskip
\noindent
{\it Comparison with non-cooperative game:}
As expected, the approach based on the pure non-cooperative game, denoted by Non-coop in
Figs.~\ref{proportion_fair_gain}-\ref{weight_sum_rate_gain},  yields inferior
performance in terms of the network utility.
The reason  is that when each BS  optimizes
only its own utility function, it does not account for the
disutility it causes to the users served by other BSs due to the interference
it generates. In economic
terms, a disutility of one agent due to the action of another is
referred to as a negative externality, which is the root of the
inefficiencies of the non-cooperative game.
For large $M$,  the system
becomes interference limited, and the
pricing mechanism in Algorithm 1 can significantly increase the
achievable network utility
by implicitly inducing cooperation and yet   maintaining the
non-cooperative nature of the  beamforming solution.

\medskip
\noindent
{\it Comparison with full-cooperation based algorithm:}
The  full-cooperation based algorithm has also been simulated. The  results show that its utility performance is always similar to that of our proposed Algorithm 1. Thus, for clarity we did not plot its utility performance in Figs.~\ref{proportion_fair_gain}-\ref{weight_sum_rate_gain}. Notice that this performance similarity is reasonable: because the optimization problem of downlink beamforming in multi-cell OFDMA networks is non-convex, even full-cooperation based algorithm only achieves the KKT solution. For our proposed Algorithm 1, the total utility at the NE point is no less than that at the point satisfying the KKT conditions of the original problem.

\medskip
\noindent
{\it Comparison with distributed interference alignment:}
The maximum SINR interference alignment (MSIA) method is an extension of
the interference alignment algorithm proposed in \cite{Gomadam2009},
where the receive filters are chosen to maximize the SINR,
and in the meantime to minimize the leakage interference
at the receivers.
For the MISO scenario discussed in this paper, interference
alignment can be accomplished through symbol extension.
In the simulations, the number of symbol extension $S$ is set from 2 to 10, and
the degree   of freedom $d$  for a user's message is set from 1 to its upper bound $S$.
In Fig. (\ref{proportion_fair_gain})-(\ref{weight_sum_rate_gain}), we plot the performance of MSIA for
the case of the best choices of symbol extension  $S$ and the degree  of freedom $d$.

It is seen that the performance of MSIA is inferior to that of Algorithm 1, especially
at the strong interference scenario. In fact, the MSIA is even inferior to the per-cell optimized
non-cooperative game solution in some scenarios.
The reasons are as follows.

\begin{itemize}
\item The MSIA only optimizes the beam directions. In contrast,
Algorithm 1 optimizes both the beam directions and the power distribution
across the beams.

\item  The MSIA is designed to maximize the SINR  and in the meantime to minimize the leakage interference.
In contrast, Algorithm 1 is designed to optimize a general utility function.

\item In MSIA, the iterative algorithm alternates between the original and
reciprocal networks. Within each network, the receivers
update their interference suppression filters. In contrast, Algorithm 1 is iteratively implemented only at the transmitters (BSs)
and it can converge  only in a small number of
iterations. Thus, Algorithm 1 incurs a much lower information exchange overhead, in comparison with MSIA.

\end{itemize}

\section{Conclusions}
We have considered the downlink beamforming problem for   co-channel interference mitigation
in multi-cell OFDMA networks.
The problem is formulated as a general utility maximization problem
 subject to the per-cell power constraints, which is non-convex.
 We have proposed a distributed solution based on the non-cooperative game with
 pricing mechanism.
 We have shown that for some popular utility functions, such as the
 weighted sum rate utility, the proportional fairness utility, and
 the $\alpha$-fairness utility, the proposed algorithm converges
 to a Nash equilibrium point.
   Moreover,
we have developed an efficient algorithm to solve the KKT condition at each base station based on the
dual decomposition technique.
We have provided extensive simulation results to illustrate that the proposed method can
converge to a Nash equilibrium in a small number of iterations, and it outperforms several
state-of-the-art approaches to multicell interference mitigation, including the recently
developed distributed interference alignment method.

In this paper, we have assumed the perfect instantaneous channel state information. From the practical point of view,   the following issues remain to be investigated in future:
(1)  the impact of the reduced information exchange;
(2)  the robustness of the proposed method in the presence of  transmission delay, packet loss and estimation error.

\section*{Appendix A: Proof of Proposition 1}
\label{sec:appendix}


After some simple manipulations, we have
\footnote{Here, we drop the explicit dependency of $U_{m,k}^{(n)}$ on $\Gamma_{m,k}^{(n)}$, and denote $U_{m,k}^{(n)}(\Gamma_{m,k}^{(n)}(\mathbf{W}^{(n)}))$ as $U_{m,k}^{(n)}(\mathbf{W}^{(n)})$.}
\begin{equation}\label{qqq}
 \frac{{\partial ^2 U_{m,k}^{^{(n)}} ({\mathbf{W}^{(n)}})}}{{\partial \left( {\mathcal{I}_{m,k}^{^{(n)}} } \right)^2 }}
  = \left( {U_{m,k}^{^{(n)}} } \right)^\prime   {\frac{{\left| {{\vec{\mathbf{h}}_{m,k}^{^{(n)}} } \mathbf{w}_{m,k}^{^{(n)}} } \right|^2 }}{{\left( {1 + \mathcal{I}_{m,k}^{^{(n)}} } \right)^3 }}} \left( {2 - \kappa_{m,k}^{^{(n)}} } \right).
\end{equation}
By the assumptions on the utility function $U_{m,k}^{^{(n)}}$, we have $\left( {U_{m,k}^{^{(n)}} } \right)^\prime>0$, and $0 \leq \kappa_{m,k}^{^{(n)}}\leq 2$. Thus
\begin{equation}
\frac{{\partial ^2 U_{m,k}^{^{(n)}} ({\mathbf{W}^{(n)}})}}{{\partial \left( {\mathcal{I}_{m,k}^{^{(n)}} } \right)^2 }} \ge 0.
\end{equation}
Hence
$U_{m,k}^{^{(n)}}$ is a convex function of  $\mathcal{I}_{m,k}^{^{(n)}}$. Then we have
\begin{equation}\label{Appedix1}
\begin{split}
 U_{m,k}^{(n)} ({\mathbf{\widehat{W}}^{(n)}})
  \ge U_{m,k}^{(n)} ({\mathbf{W}^{(n)}}) + \frac{{\partial U_{m,k}^{(n)} }}{{\partial  \mathcal{I}_{m,k}^{(n)} }}\left| {_{\mathcal{I}_{m,k}^{(n)} } } \right.\left( { \mathcal{\widehat{I}}_{m,k}^{(n)}  - \mathcal{I}_{m,k}^{(n)} } \right) \\
  = U_{m,k}^{(n)} ({\mathbf{W}^{(n)}}) - \pi _{m,k}^{(n)} ({\mathbf{W}^{(n)}} )\left( {\mathcal{\widehat{I}}_{m,k}^{(n)}  -  \mathcal{I}_{m,k}^{(n)} } \right),
  \end{split}
\end{equation}
where  $\mathcal{I}_{m,k}^{^{(n)}}$ and $\mathcal{\widehat{I}}_{m,k}^{^{(n)}}$ denote the interferences at the current operating point $\mathbf{W}=\{\mathbf{W}^{(n)}, n \in \mathcal{N}\}$ and at any new operating point  $\mathbf{\widehat{W}}=\{\mathbf{\widehat{W}}^{(n)}, n \in \mathcal{N}\}$, respectively.

Summing up (\ref{Appedix1}) over all users served by BS $m$ and over all sub-channels, we have
(\ref{Appedix2}).
\begin{figure*}[!t]
\begin{equation}\label{Appedix2}
\sum\limits_{n \in \mathcal{N}} {\sum\limits_{k \in \mathcal{B}_m ^{(n)}} {U_{m,k}^{(n)} ({\mathbf{\widehat{W}}^{(n)}})} } {\rm{ }}
  \ge \sum\limits_{n \in \mathcal{N}} {\sum\limits_{k \in \mathcal{B}_m ^{(n)}} {\left[ {U_{m,k}^{(n)} ({\mathbf{W}^{(n)}}) - \pi _{m,k}^{(n)} ({\mathbf{W}^{(n)}})\left( { \mathcal{\widehat{I}}_{m,k}^{(n)}  -  \mathcal{I}_{{\rm{ }}m,k}^{(n)} } \right)} \right]} }.
\end{equation}
\end{figure*}

Hereafter we will drop the explicit dependency of $\pi _{m,k}^{^{(n)}}$ on $\mathbf{W}^{(n)}$.

Assume that
 BS $m$ applies Algorithm 1 to update its beam-vectors, given the current beam-vectors $\mathbf{W}=\{\mathbf{W}^{(n)}, n \in \mathcal{N}\}$.
According to the condition of STEP 6 in Algorithm 1, we have
\begin{equation}\label{Appedix3A}
\bar{U}_m(\bar{\mathbf{W}}_m,\mathbf{W}_{-m}) \geq
  \bar{U}_m(\mathbf{W}_m,\mathbf{W}_{-m}),
\end{equation}
where $
{\mathbf{\bar W}} = \left\{ {{\mathbf{W}}_1 , \cdots ,{\mathbf{W}}_{m - 1} ,{\mathbf{\bar W}}_m ,{\mathbf{W}}_{m + 1} , \cdots ,{\mathbf{W}}_M } \right\}
$  denotes the operating point after BS $m$ updates its beam-vectors ${\mathbf{\bar W}}_m=\{\mathbf{\bar{w}}_{m,k}^{(n)}, k \in \mathcal{B}_{m}^{(n)}, n \in \mathcal{N}\}$.

Plugging (\ref{Equ7 8})-(\ref{totalprice2}) into (\ref{Appedix3A}),  we have (\ref{Appedix3}).
\begin{figure*}[!t]
\begin{equation}\label{Appedix3}
\begin{split}
 &\hspace{0.5cm}{\rm{   }}\sum\limits_{n \in \mathcal{N}} {\sum\limits_{k \in \mathcal{B}_m ^{(n)}} {\Big( {U_{m,k}^{^{(n)}} \left( {{\mathbf{\bar W}^{(n)}}} \right){\rm{ - }}\sum\limits_{j \in \mathcal{M}} {\sum\limits_{\scriptstyle \;\;\; u \in \mathcal{B}_j ^{(n)} \hfill \atop
  \scriptstyle (j,u) \ne (m,k) \hfill} {\pi _{j,u}^{^{(n)}} \left| {{\vec{\mathbf{h}}_{m,u}^{^{(n)}} }\mathbf{\bar{w}}_{m,k}^{^{(n)}} } \right|^2 } } } \Big)} }\\
   &\ge \sum\limits_{n \in \mathcal{N}} {\sum\limits_{k \in \mathcal{B}_m ^{(n)}} {\Big( {U_{m,k}^{^{(n)}} \left( {\mathbf{W}^{(n)}} \right){\rm{ - }}\sum\limits_{j \in \mathcal{M}} {\sum\limits_{\scriptstyle \;\;\; u \in \mathcal{B}_j ^{(n)} \hfill \atop
  \scriptstyle (j,u) \ne (m,k) \hfill} {\pi _{j,u}^{^{(n)}} \left| { {\vec{\mathbf{h}}_{m,u}^{^{(n)}} }  \mathbf{w}_{m,k}^{^{(n)}} } \right|^2 } } } \Big)} } {\rm{ }}.
 \end{split}
\end{equation}
\end{figure*}

Because of BS $m$'s beam-vector update, the received interference by user
$u \in \mathcal{B}_j ^{(n)},j \in \mathcal{M} \backslash m$ is changed from $\mathcal{I}_{j,u}^{^{(n)}}$ to  $\mathcal{\bar{I}}_{j,u}^{^{(n)}}$, and
\begin{equation}\label{Appedix4}
\mathcal{\bar{I}}_{j,u}^{^{(n)}}  - \mathcal{I}_{j,u}^{^{(n)}} {\rm{  }}
   = \sum\limits_{k \in \mathcal{B}_m ^{(n)}} {\Big( {\Big| {{\vec{\mathbf{h}}_{m,u}^{^{(n)}} }\mathbf{\bar{w}}_{m,k}^{^{(n)}} } \Big|^2  - \Big| { {\vec{\mathbf{h}}_{m,u}^{^{(n)}} }  \mathbf{w}_{m,k}^{^{(n)}} } \Big|^2 } \Big)}.
\end{equation}
Thus, we have (\ref{Appedix5}).
\begin{figure*}[!t]
\begin{equation}\label{Appedix5}
{\rm{  }}\sum\limits_{\scriptstyle j \in \mathcal{M} \hfill \atop
  \scriptstyle j \ne m \hfill} { {\sum\limits_{n \in \mathcal{N}} {\sum\limits_{u \in \mathcal{B}_j ^{(n)}} { {\pi _{j,u}^{(n)} \Big( {\mathcal{\bar{I}}_{j,u}^{(n)}  - \mathcal{I}_{j,u}^{(n)} } } \Big)} } } } 
  = \sum\limits_{n \in \mathcal{N}} {\sum\limits_{k \in \mathcal{B}_m ^{(n)}} { {\sum\limits_{j \in \mathcal{M}} {\sum\limits_{\;\;\;\; \scriptstyle u \in \mathcal{B}_j ^{(n)} \hfill \atop
  \scriptstyle (j,u) \ne (m,k) \hfill} {\pi _{j,u}^{(n)} \Big( {\Big| {{\vec{\mathbf{h}}_{m,u}^{(n)} }  \mathbf{\bar{w}}_{m,k}^{(n)} } \Big|^2  -
   \Big| { {\vec{\mathbf{h}}_{m,u}^{(n)} } \mathbf{w}_{m,k}^{(n)} } \Big|^2 } \Big)} } } } }.
\end{equation}
\end{figure*}

Adding the constant terms
$
\sum\limits_{\scriptstyle j \in \mathcal{M} \hfill \atop
  \scriptstyle j \ne m \hfill} {\Big[ {\sum\limits_{n \in \mathcal{N}} {\sum\limits_{u \in \mathcal{B}_j ^{(n)}} {U_{j,u}^{^{(n)}} ({\mathbf{W}^{(n)}})} } } \Big]}
$
to both sides of (\ref{Appedix3}), we have (\ref{Appedix6}).
\begin{figure*}[!t]
\begin{equation}\label{Appedix6}
\begin{split}
 &\hspace{0.5cm}{\rm{   }}\sum\limits_{n \in \mathcal{N}} {\sum\limits_{k \in \mathcal{B}_m ^{(n)}} {\Big[ {U_{m,k}^{^{(n)}} \left( {{\mathbf{\bar W}^{(n)}}} \right){\rm{ - }}\sum\limits_{j \in \mathcal{M}} {\sum\limits_{\;\;\;\; \scriptstyle u \in \mathcal{B}_j ^{(n)} \hfill \atop
  \scriptstyle (j,u) \ne (m,k) \hfill} {\pi _{j,u}^{^{(n)} } \left| {{\vec{\mathbf{h}}_{m,u}^{^{(n)}} }\mathbf{\bar{w}}_{m,k}^{^{(n)}} } \right|^2 } } } \Big]} }  + \sum\limits_{\scriptstyle j \in \mathcal{M} \hfill \atop
  \scriptstyle j \ne m \hfill} { {\sum\limits_{n \in \mathcal{N}} {\sum\limits_{u \in \mathcal{B}_j ^{(n)}} {U_{j,u}^{^{(n)}} ({\mathbf{W}^{(n)}})} } } }  \\
  &= \sum\limits_{n \in \mathcal{N}} {\sum\limits_{k \in \mathcal{B}_m ^{(n)}} { {U_{m,k}^{^{(n)}} \left( {{\mathbf{\bar W}^{(n)}}} \right)} } }  + \sum\limits_{\scriptstyle j \in \mathcal{M} \hfill \atop
  \scriptstyle j \ne m \hfill} { {\sum\limits_{n \in \mathcal{N}} {\sum\limits_{u \in \mathcal{B}_j ^{(n)}} {U_{j,u}^{^{(n)}} ({\mathbf{W}^{(n)}})} } } } {\rm{ - }}\sum\limits_{n \in \mathcal{N}} {\sum\limits_{k \in \mathcal{B}_m ^{(n)}} { {\sum\limits_{j \in \mathcal{M}} {\sum\limits_{\;\;\;\; \scriptstyle u \in \mathcal{B}_j ^{(n)} \hfill \atop
  \scriptstyle (j,u) \ne (m,k) \hfill} {\pi _{j,u}^{^{(n)}} \left| {{\vec{\mathbf{h}}_{m,u}^{^{(n)}} } \mathbf{\bar{w}}_{m,k}^{^{(n)}} } \right|^2 } } } } }
  \\
  &\ge \sum\limits_{j \in \mathcal{M}} { {\sum\limits_{n \in \mathcal{N}} {\sum\limits_{u \in \mathcal{B}_j ^{(n)}} {U_{j,u}^{^{(n)}} ({\mathbf{W}^{(n)}})} } } } {\rm{ - }}\sum\limits_{n \in \mathcal{N}} {\sum\limits_{k \in \mathcal{B}_m ^{(n)}} { {\sum\limits_{j \in \mathcal{M}} {\sum\limits_{\;\;\;\; \scriptstyle u \in \mathcal{B}_j ^{(n)} \hfill \atop
  \scriptstyle (j,u) \ne (m,k) \hfill} {\pi _{j,u}^{^{(n)}} \left| {{\vec{\mathbf{h}}_{m,u}^{^{(n)}} }  \mathbf{w}_{m,k}^{^{(n)}} } \right|^2 } } } } }.
 \end{split}
\end{equation}
\end{figure*}

Subtracting the terms\\
$
\sum\limits_{n \in \mathcal{N}} {\sum\limits_{k \in \mathcal{B}_m ^{(n)}} { {\sum\limits_{j \in \mathcal{M}} {\sum\limits_{\;\;\;\; \scriptstyle u \in \mathcal{B}_j ^{(n)} \hfill \atop
  \scriptstyle (j,u) \ne (m,k) \hfill} {\pi _{j,u}^{^{(n)}} \left| {{\vec{\mathbf{h}}_{m,u}^{^{(n)}} }  \mathbf{w}_{m,k}^{^{(n)}} } \right|^2 } } } } }
$
from both sides of (\ref{Appedix6}), and using (\ref{Appedix5}), we have (\ref{Appedix7}).
\begin{figure*}[!t]
\begin{equation}\label{Appedix7}
\begin{split}
&\;\;\;\; {\rm{  }}\sum\limits_{n \in \mathcal{N}} {\sum\limits_{k \in \mathcal{B}_m ^{(n)}} {  {U_{m,k}^{^{(n)}} \left( {{\mathbf{\bar W}^{(n)}}} \right)} } }  + \sum\limits_{\scriptstyle j \in \mathcal{M} \hfill \atop
 \scriptstyle j \ne m \hfill} { {\sum\limits_{n \in \mathcal{N}} {\sum\limits_{u \in \mathcal{B}_j ^{(n)}} {U_{j,u}^{^{(n)}} ({\mathbf{W}^{(n)}})} } } }
 \;\;\;\;\;\;\;\; \hspace{-0.7cm} {\rm{ - }}\sum\limits_{n \in \mathcal{N}} {\sum\limits_{k \in \mathcal{B}_m ^{(n)}} { {\sum\limits_{j \in \mathcal{M}} {\sum\limits_{\;\;\;\; \scriptstyle u \in \mathcal{B}_j ^{(n)} \hfill \atop
  \scriptstyle (j,u) \ne (m,k) \hfill} {\pi _{j,u}^{^{(n)}} \Big( {\left| {{\vec{\mathbf{h}}_{m,u}^{^{(n)}} } \mathbf{\bar{w}}_{m,k}^{^{(n)}} } \right|^2  - \left| {{\vec{\mathbf{h}}_{m,u}^{^{(n)}} } \mathbf{w}_{m,k}^{^{(n)}} } \right|^2 } \Big)} } } } }  \\
  &= \sum\limits_{n \in \mathcal{N}} {\sum\limits_{k \in \mathcal{B}_m ^{(n)}} {  {U_{m,k}^{^{(n)}} \left( {{\mathbf{\bar W}^{(n)}}} \right)}  } }  + \sum\limits_{\scriptstyle j \in \mathcal{M} \hfill \atop
  \scriptstyle j \ne m \hfill} {  {\sum\limits_{n \in \mathcal{N}} {\sum\limits_{u \in \mathcal{B}_j ^{(n)}} {U_{j,u}^{^{(n)}} ({\mathbf{W}^{(n)}})} } }  }
   \;\;\;\;\;\;\;\; \hspace{-0.7cm}- \sum\limits_{\scriptstyle j \in \mathcal{M} \hfill \atop
  \scriptstyle j \ne m \hfill} {  {\sum\limits_{n \in \mathcal{N}} {\sum\limits_{u \in \mathcal{B}_j ^{(n)}}    \pi _{j,u}^{(n)}  \left( {\mathcal{\bar{I}}_{j,u}^{^{(n)}}  - \mathcal{I}_{j,u}^{^{(n)}} } \right)}  }  }  \\
 &= \sum_{n \in \mathcal{N}}  \sum_{k \in \mathcal{B}_m^{(n)} }    U_{m,k}^{(n)}
 \left(  \bar{\mathbf{W}}^{(n)} \right)
 + \sum_{\scriptstyle j \in \mathcal{M} \hfill \atop
   \scriptstyle j \ne m \hfill}    {\sum_{n \in \mathcal{N}} {\sum_{u \in \mathcal{B}_j ^{(n)}} {\left[ {U_{j,u}^{^{(n)}} ({\mathbf{W}^{(n)}}) - \pi _{j,u}^{^{(n)}} \left( {\mathcal{\bar{I}}_{j,u}^{^{(n)}}  - \mathcal{I}_{j,u}^{^{(n)}} } \right)} \right]} } }    \ge \sum_{j \in \mathcal{M}}\sum_{n \in \mathcal{N}}\sum_{u \in \mathcal{B}_j ^{(n)}}  U_{j,u}^{(n)}  ({\mathbf{W}}^{(n)}).
 \end{split}
 \end{equation}
 \hrulefill
 \end{figure*}

Summing up  (\ref{Appedix2}) over
$j \in \mathcal{M}\backslash m$
at the updated operating point $\mathbf{\bar{W}}$, we have
\begin{eqnarray}\label{Appedix8}
&&\hspace{-1.0cm}{\rm{     }}\sum\limits_{\scriptstyle j \in \mathcal{M} \hfill \atop
  \scriptstyle j \ne m \hfill} {  {\sum\limits_{n \in \mathcal{N}} {\sum\limits_{u \in \mathcal{B}_j ^{(n)}} {U_{j,u}^{^{(n)}} ({\mathbf{\bar{W}}^{(n)}})} } }  } {\rm{ }}.\nonumber \\
 &&\hspace{-1.5cm}\ge \sum\limits_{\scriptstyle j \in \mathcal{M} \hfill \atop
  \scriptstyle j \ne m \hfill} {  {\sum\limits_{n \in \mathcal{N}} {\sum\limits_{u \in \mathcal{B}_j ^{(n)}} {\left[ {U_{j,u}^{^{(n)}} ({\mathbf{W}^{(n)}}) - \pi _{j,u}^{^{(n)}} \left( {\mathcal{\bar{I}}_{j,u}^{^{(n)}}  - \mathcal{I}_{j,u}^{^{(n)}} } \right)} \right]} } }  }
\end{eqnarray}

Adding $\sum\limits_{n \in \mathcal{N}} {\sum\limits_{k \in \mathcal{B}_m ^{(n)}} {U_{m,k}^{^{(n)}} ({\mathbf{\bar{W}}^{(n)}})} }$
to both sides of (\ref{Appedix8}), we have (\ref{Appedix9}).
\begin{figure*}[!t]
\begin{equation}\label{Appedix9}
\begin{split}
 \sum\limits_{j \in \mathcal{M}} { {\sum\limits_{n \in \mathcal{N}} {\sum\limits_{u \in \mathcal{B}_j ^{(n)}} {U_{j,u}^{^{(n)}} ({\mathbf{\bar W}^{(n)}})} } } }
  \ge \sum\limits_{n \in \mathcal{N}} {\sum\limits_{k \in \mathcal{B}_m ^{(n)}} {U_{m,k}^{^{(n)}} ({\mathbf{\bar W}^{(n)}})} }  + \sum\limits_{\scriptstyle j \in \mathcal{M} \hfill \atop
  \scriptstyle j \ne m \hfill} {  {\sum\limits_{n \in \mathcal{N}} {\sum\limits_{u \in \mathcal{B}_j ^{(n)}} {\left[ {U_{j,u}^{^{(n)}} ({\mathbf{W}^{(n)}}) - \pi _{j,u}^{^{(n)}} \left( {\mathcal{\bar{I}}_{j,u}^{^{(n)}}  - \mathcal{I}_{{\rm{   }}j,u}^{^{(n)}} } \right)} \right]} } } }.
 \end{split}
\end{equation}
\hrulefill
\end{figure*}

Combining  (\ref{Appedix7}) and  (\ref{Appedix9}) yields
\begin{equation}\label{Appedix10}
\underbrace{\sum\limits_{j \in \mathcal{M}} {  {\sum\limits_{n \in \mathcal{N}} {\sum\limits_{u \in \mathcal{B}_j ^{(n)}} {U_{j,u}^{^{(n)}} ({\mathbf{\bar W}^{(n)}})} } } } }_{ U_{\rm network} ({\mathbf{\bar W}}) } \ge
\underbrace{ \sum\limits_{j \in \mathcal{M}} {  {\sum\limits_{n \in \mathcal{N}} {\sum\limits_{u \in \mathcal{B}_j ^{(n)}} {U_{j,u}^{^{(n)}} ({\mathbf{W}^{(n)}})} } }  } }_{U_{\rm network} ({\mathbf{W}})}.
\end{equation}

Hence, when BS $m$ adjusts its beam-vectors,   the total utility cannot decrease.
Since at most one BS updates its beam-vectors  at anytime, the total utility is non-decreasing in each iteration.
As both the number of players and the size of
the  strategy sets are finite, the total utility is bounded. Thus, the total utility will convergence,

Now, we assume that Algorithm 1 converges to a fixed
point   $\mathbf{W}^{*}=(\mathbf{W}_1^{*},\ldots,\mathbf{W}_M^{*})$.
If $\mathbf{W}^{*}$ is not an NE point, then
  there exists   $\mathbf{\widetilde{W}}=(\mathbf{W}_1^{*},\ldots, \mathbf{\widetilde{W}}_m,\ldots,\mathbf{W}_M^{*})$, such that
\begin{equation}
\bar{U}_m(\mathbf{\widetilde{W}}_m,\mathbf{W}_{-m}^{*}) \geq
  \bar{U}_m(\mathbf{W}_m^{*},\mathbf{W}_{-m}^{*}).
\end{equation}
Applying the similar deduction in (\ref{Appedix3A})-(\ref{Appedix10}), we have
\begin{equation}\label{Appedix12}
U_{\rm network} ({\mathbf{\widetilde{W}}})  \ge U_{\rm network} ({\mathbf{W}^{*}}),
\end{equation}
which contradicts the assumption that $\mathbf{W}^{*}$ is a fixed point.
Thus  $\mathbf{W}^{*}$ is an NE point. \hfill $\Box$

\section*{Appendix B: Proof of Proposition 2}

The proof is along the similar line of that for Proposition 1 in \cite{venturino:2010}.
There are two cases for the solution to the KKT conditions of  (\ref{BSsubproblem}).

\noindent
\textit{Case 1}: $\lambda _m > 0$ and $\mathbf{w}_{m,k}^{(n)}\neq \mathbf{0}$.

\noindent
Notice that $\mathbf{T}_{m,k}^{(n)}$  is a positive-definite
matrix. Thus, it is easy to have
\begin{equation} \label{ss1}
 \mathbf{T}_{m,k}^{(n)}\mathbf{w}_{m,k}^{(n)} \neq \mathbf{0}.
\end{equation}
Obviously, (\ref{ss1}) and (\ref{KKT}) imply
that $\vec {\mathbf{h}}_{m,k}^{(n)} \mathbf{w}_{m,k}^{(n)} \neq \mathbf{0}$
and $\mathbf{h}_{m,k}^{(n)} \propto \mathbf{T}_{m,k}^{(n)}\mathbf{w}_{m,k}^{(n)}$.
Hence, a non-zero solution $\mathbf{w}_{m,k}^{(n)*}$ to the KKT conditions of (\ref{BSsubproblem})
must be of the form
\begin{equation}\label{wop1}
 \mathbf{w}_{m,k}^{(n)*} \propto \mathbf{T}_{m,k}^{(n)\dag}\mathbf{h}_{m,k}^{(n)}.
\end{equation}

\noindent
\textit{Case 2}: $\lambda _m = 0$ and $\mathbf{w}_{m,k}^{(n)}\neq \mathbf{0}$.

\noindent
It is easy to see that
(\ref{KKT}) is satisfied only if one of the following two conditions
holds:

(a)  $\vec {\mathbf{h}}_{m,k}^{(n)} \mathbf{w}_{m,k}^{(n)} = 0$ and $\mathbf{L}_{m,k}^{(n)}\mathbf{w}_{m,k}^{(n)} = 0$;

(b) $\mathbf{h}_{m,k}^{(n)} \in \varrho(\mathbf{L}_{m,k}^{(n)})$.

If (a) holds, $\vec {\mathbf{h}}_{m,k}^{(n)} \mathbf{w}_{m,k}^{(n)} = 0$  implies that the non-zero beam-vector $\mathbf{w}_{m,k}^{(n)}$
is orthogonal to the channel vector $\vec {\mathbf{h}}_{m,k}^{(n)}$. In this case, user $k \in \mathcal{B}_{m}^{(n)}$  cannot receive
any information from the
serving base station. Hence
we discard the solutions
$\vec {\mathbf{h}}_{m,k}^{(n)} \mathbf{w}_{m,k}^{(n)} = 0$ and $\mathbf{L}_{m,k}^{(n)}\mathbf{w}_{m,k}^{(n)} = \mathbf{0}$
for the case $\lambda _m \neq 0$ and $\mathbf{w}_{m,k}^{(n)}\neq \mathbf{0}$.

If (b) holds and $\lambda _m = 0$, a non-zero beam-vector
which satisfies (\ref{KKT})  must be of the form
\begin{equation}\label{wop2}
 \mathbf{w}_{m,k}^{(n)*} \propto \mathbf{L}_{m,k}^{(n)\dag}\mathbf{h}_{m,k}^{(n)}.
\end{equation}
Notice that  (\ref{wop2}) is equivalent to (\ref{wop1}) with $\lambda _m = 0$.
Meanwhile, notice
that multiplying $\mathbf{w}_{m,k}^{(n)*}$ by any unit-norm complex number
does not affect either the objective function  or
the power constraint in  (\ref{BSsubproblem}).
Hence, for the two cases discussed above,
we can set the unique solution   to the KKT conditions of   (\ref{BSsubproblem}) as
\begin{equation}\label{beamsolution}
 \mathbf{w}_{m,k}^{(n) * }  = \beta_{m,k}^{(n)}\mathbf{T}_{m,k}^{(n) \dag }  \mathbf{h}_{m,k}^{(n)},
\end{equation}
where $\beta_{m,k}^{(n)}$ is some scalar constant.

To determine $\beta_{m,k}^{(n)}$, we plug
 (\ref{beamsolution}) into (\ref{KKT}) to obtain
\begin{eqnarray}\label{Appedix8}
& {U_{m,k}^{(n)} \Big( {\frac{{\left| {\vec {\mathbf{h}}_{m,k}^{(n)} \beta _{m,k}^{(n)} \mathbf{T}_{m,k}^{(n)\dag} \mathbf{h}_{m,k}^{(n)} } \right|^2 }}{{1 + \mathcal{I}_{m,k}^{(n)} }}} \Big)}^\prime  \frac{{\mathbf{h}_{m,k}^{(n)} \vec {\mathbf{h}}_{m,k}^{(n)} \beta _{m,k}^{(n)} \mathbf{T}_{m,k}^{(n)\dag}   \mathbf{h}_{m,k}^{(n)} }}{{1 + \mathcal{I}_{m,k}^{(n)} }} \nonumber \\
 &\hspace{-5cm} = \mathbf{T}_{m,k}^{(n)} \beta _{m,k}^{(n)} \mathbf{T}_{m,k}^{(n)\dag}  \mathbf{h}_{m,k}^{(n)}
\end{eqnarray}

Considering $\mathbf{T}_{m,k}^{(n)} \beta _{m,k}^{(n)} \mathbf{T}_{m,k}^{(n)\dag}=\beta _{m,k}^{(n)}\mathbf{I}$, and $\mathbf{h}_{m,k}^{(n)}\neq \mathbf{0}$, we have
\begin{equation}
  {U_{m,k}^{(n)} \Big( {\frac{{| {\vec {\mathbf{h}}_{m,k}^{(n)} \beta _{m,k}^{(n)} \mathbf{T}_{m,k}^{(n) \dag }  \mathbf{h}_{m,k}^{(n)} } |^2 }}{{1 + \mathcal{I}_{m,k}^{(n)} }}} \Big)}^\prime   = \frac{{1 + \mathcal{I}_{m,k}^{(n)} }}{{\vec {\mathbf{h}}_{m,k}^{(n)} \mathbf{T}_{m,k}^{(n) \dag }  \mathbf{h}_{m,k}^{(n)} }} \\
\end{equation}
which is equivalent to
\begin{equation}
\frac{{\left| {\vec {\mathbf{h}}_{m,k}^{(n)} \beta _{m,k}^{(n)} \mathbf{T}_{m,k}^{(n) \dag }  \mathbf{h}_{m,k}^{(n)} } \right|^2 }}{{1 + \mathcal{I}_{m,k}^{(n)} }} =  {\rm Inv} \Big\{ U_{m,k}^{(n)} \Big( {\frac{{1 + \mathcal{I}_{m,k}^{(n)} }}{{\vec {\mathbf{h}}_{m,k}^{(n)} \mathbf{T}_{m,k}^{(n) \dag }  \mathbf{h}_{m,k}^{(n)} }}} \Big)^\prime \Big\}.
\end{equation}
Thus, we obtain
\begin{equation}\label{Beta}
\beta _{m,k}^{(n)}  = \sqrt {\left( {1 + \mathcal{I}_{m,k}^{(n)} } \right)\Phi _{m,k}^{(n)} \Upsilon _{m,k}^{(n)} }.
\end{equation}
 \hfill $\Box$

\bibliographystyle{ieeetr}

\begin{thebibliography}{10}

\bibitem{Gesbert2010JSAC}
D.~Gesbert, S.~Hanly, H.~Huang, S.~Shamai, O.~Simeone, and W.~Yu, ``{Multi-cell
  MIMO cooperative networks: A new look at interference},'' {\em IEEE J.
  Select. Areas Commun.}, vol.~28, no.~9, pp.~1--29, 2010.

\bibitem{Zhang:2004}
H.~Zhang and H.~Dai, ``Cochannel interference mitigation and cooperative
  processing in downlink multicell multiuser mimo networks,'' {\em EURASIP J.
  Wireless Commun. Netw.}, vol.~2004, no.~2, pp.~222--235, 2004.

\bibitem{Ng:2008}
B.~Ng, J.~Evans, S.~Hanly, and D.~Aktas, ``Distributed downlink beamforming
  with cooperative base stations,'' {\em IEEE Tran. Inform. Theory}, vol.~54,
  no.~12, pp.~5491--5499, 2008.

\bibitem{Somekh:2007}
B.~Somekh, O.and~Zaidel and S.~Shamai, ``Sum rate characterization of joint
  multiple cell-site processing,'' {\em IEEE Trans. Inform. Theory}, vol.~53,
  no.~12, pp.~4473--4497,  2007.

\bibitem{Shamai:2007}
S.~Shamai, O.~S. A. Z. B.~M. Somekh, O.and~Simeone, and H.~V. Poor,
  ``Cooperative multi-cell networks: Impact of limited-capacity backhaul and
  inter-users links,'' in {\em  Proc.  Joint Workshop on Coding and
  Communications}, Durnstein, Austria, pp.~14--16, 2007.

\bibitem{Simeone:2009}
O.~Simeone, O.~Somekh, H.~V. Poor, and S.~Shamai, ``Local base station
  cooperation via finite-capacity links for the uplink of linear cellular
  networks,'' {\em IEEE Trans. Inform. Theory}, vol.~55, no.~1,
  pp.~190--204, 2009.

\bibitem{Tamaki:2007}
T.~Tamaki, K.~Seong, and J.~Cioffi, ``Downlink {MIMO} systems using cooperation
  among base stations in a slow fading channel,'' {\em
  Proc. 2007 IEEE Int. Conf. Commun. (ICC'07)},  pp.~4728--4733, 2007.



\bibitem{Marsch:2009}
P.~Marsch and G.~Fettweis, ``On downlink network mimo under a constrained
  backhaul and imperfect channel knowledge,'' {\em Proc. 2009 IEEE Global
  Telecommun. Conf. (Globecom'09)}, 2009.

\bibitem{MAwad2010}
M. Awad, V. Mahinthan, M. Mehrjoo, X. Shen, and J.W. Mark,  ``A Dual Decomposition-based Resource Allocation for OFDMA Networks with Imperfect CSI,'' {\em IEEE Trans. Veh. Commun.}, vol. ~59, no. ~5, pp. ~2394--2403, 2010.

\bibitem{Papadogiannis:2009}
A.~Papadogiannis, E.~Hardouin, and D.~Gesbert, ``Decentralising multi-cell
  cooperative processing on the downlink: a novel robust framework,'' {\em
  EURASIP J. Wireless Commun. Netw.}, August 2009.





\bibitem{Dahrouj:2010A}
H.~Dahrouj and W.~Yu, ``Coordinated beamforming for the multicell
  multi-antenna wireless systems,''  {\em IEEE Trans.  Wireless
  Commun.}, vol.~9, pp.~1748--1759, 2010.

\bibitem{Zakhour:2010D}
R.~Zakhour and D.~Gesbert, ``Distributed multicell MISO precoding using the
  layered virtual SINR framework,'' {\em IEEE Trans. Wireless
  Commun.}, vol.~9, pp.~2444--2448, 2010.




\bibitem{venturino:2010}
L.~Venturino, N.~Prasad, and X.~Wang, ``Coordinated linear beamforming in
  downlink multi-cell wireless networks,'' {\em IEEE Trans. Wireless
  Commun.}, vol.~9, pp.~1451--1461,  2010.

\bibitem{Zhang}
R.~Zhang and S.~Cui, ``Cooperative interference management with MISO
  beamforming,'' {\em IEEE Trans.  Sig. Proc.}, vol.~58,
  no.~10, pp.~5450--5458,  2010.




\bibitem{Jorswieck2009Magazine}
E.~Jorswieck, E.~Larsson, M.~Luise, and H.~Poor, ``{Game theory in signal
  processing and communications},'' {\em IEEE Sig.   Proc. Mag.},
  vol.~26, no.~5, pp.~17--132,   2009.

\bibitem{Chen:2011AC}
  J. Chen, Q. Yu, P. Cheng, Y. Sun, Y. Fan, and X. Shen, ``Game Theoretical Approach for Channel Allocation in Wireless Sensor and Actuator Networks,'' DOI: 10.1109/TAC.2011.2164014, {\em  to appear in IEEE Trans. Autom. Control}, 2011.

\bibitem{Yu:2010INFOCOM}
Q. Yu, J. Chen, Y. Fan, X. Shen and Y. Sun, ``Multi-Channel Assignment in Wireless Sensor Networks: A Game Theoretic Approach,'' {\em Proceedings of IEEE INFOCOM}, San Diego, Ca, USA, March 15-19, 2010.

\bibitem{QZhang}
X. Xiao, Q. Zhang, Y. Shi, and Y. Gao,  ``How Much to Share: A Repeated Game Model for Peer-to-Peer Streaming under Service Differentiation Incentives,'' {\em to appear in IEEE Trans. Parallel Distrib. Syst.}, 2011.

\bibitem{ZhuHan}
Z. Han, D. Niyato, W. Saad, T. Basar, and A. Hjorungnes, ``Game Theory in Wireless and Communication Networks: Theory, Models and Applications,'' {\em Cambridge University Press}, UK, 2011.

\bibitem{Mo:2000}
J.~Mo, J.and~Walrand, ``{Fair end-to-end window-based congestion control},''
  {\em IEEE/ACM Trans.  Netw.}, vol.~8, no.~5, pp.~556--567, 2000.

\bibitem{Gomadam2009}
K.~Gomadam, V.~R. Cadambe, and S.~A. Jafar, ``{Approaching the capacity of
  wireless networks through distributed interference alignment},'' {\em arXiv:
  0803.3816, e-print}.


\bibitem{Larsson:2008}
E.~G. Larsson and E.~Jorswieck, ``Competition versus cooperation on the miso
  interference channel,'' {\em IEEE J. Select. Areas Commun.}, vol.~26, no.~9,
  pp.~1059--1069, Sep. 2008.


\bibitem{Jorswieck:2008}
E.~A. Jorswieck, E.~G. Larsson, and D.~Danev, ``Complete characterization of
  the Pareto boundary for the MISO interference channel,'' {\em IEEE
  Trans. Sig. Proc.}, vol.~56, no.~10-2, pp.~5292--5296, 2008.

\bibitem{Saraydar:2002}
C.~Saraydar, N.~Mandayam, and D.~Goodman, ``Efficient power control via pricing
  in wireless data networks,'' {\em IEEE Tran. Commun.}, vol.~50, no.~2,
  pp.~291--303, 2002.

\bibitem{Huang:2006}
J.~Huang, R.~A. Berry, and M.~L. Honig, ``{Distributed interference
  compensation for wireless networks},'' {\em IEEE J. Select. Areas Commun.},
  vol.~24, pp.~1074--1084, May 2006.

\bibitem{Schmidt:2009}
D.~A. Schmidt, C.~Shi, R.~A. Berry, M.~L. Honig, and W.~Utschick,
  ``Distributed resource allocation schemes: Pricing algorithms for power
  control and beamformer design in interference networks,'' {\em IEEE Sig.
  Proc. Mag.},
  vol.~26, no.~5, pp.~53--63,  2009.
\end{thebibliography}

\begin{IEEEbiography}[{\includegraphics[width=1in,height=1.25in,clip,keepaspectratio]
{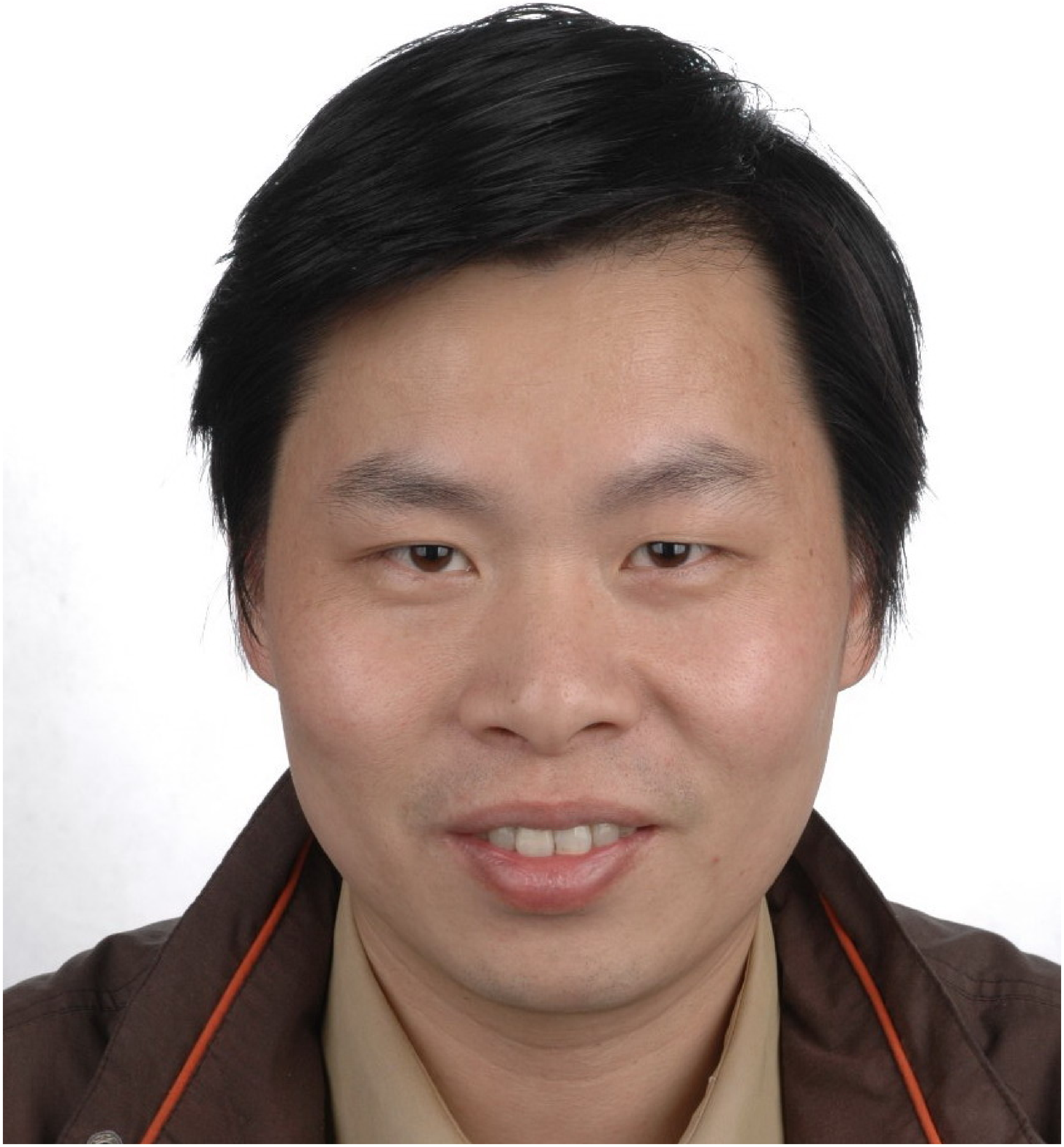}}]{Weiqiang Xu} (M'09)
 received his M.Sc. degree in Communications and Information System from Southwest Jiao-Tong University, China, and his Ph.D. degree in Control Science and Engineering from Zhejiang University, China, in 2003 and 2006, respectively. He also was a postdoctor research fellow with the group of Networked Sensing and Control in the State Key laboratory of Industrial Control Technology, Zhejiang University, China. From Oct. 2009 to Oct. 2010, he visited Prof. Xiaodong Wang's research group in Electrical Engineering Department at Columbia University, New York. He is currently a professor with the School of Information Science and Technology, Zhejiang Sci-Tech University, Hangzhou, China. His research interests include multi-cell networks, Ad Hoc networks, wireless sensor networks, wireless optical networks, congestion control, and networked control system, etc. He has served as a TPC member for IEEE Globecom 2012, IWCMC 2009, IWCMC 2010, PMSN 2009, IHMSC 2009, IHMSC 2010, IHMSC 2011, IHMSC 2012. He has also served as a peer reviewer for a variety of IEEE journals and conferences.
\end{IEEEbiography}

\begin{IEEEbiography}[{\includegraphics[width=1in,height=1.25in,clip,keepaspectratio]{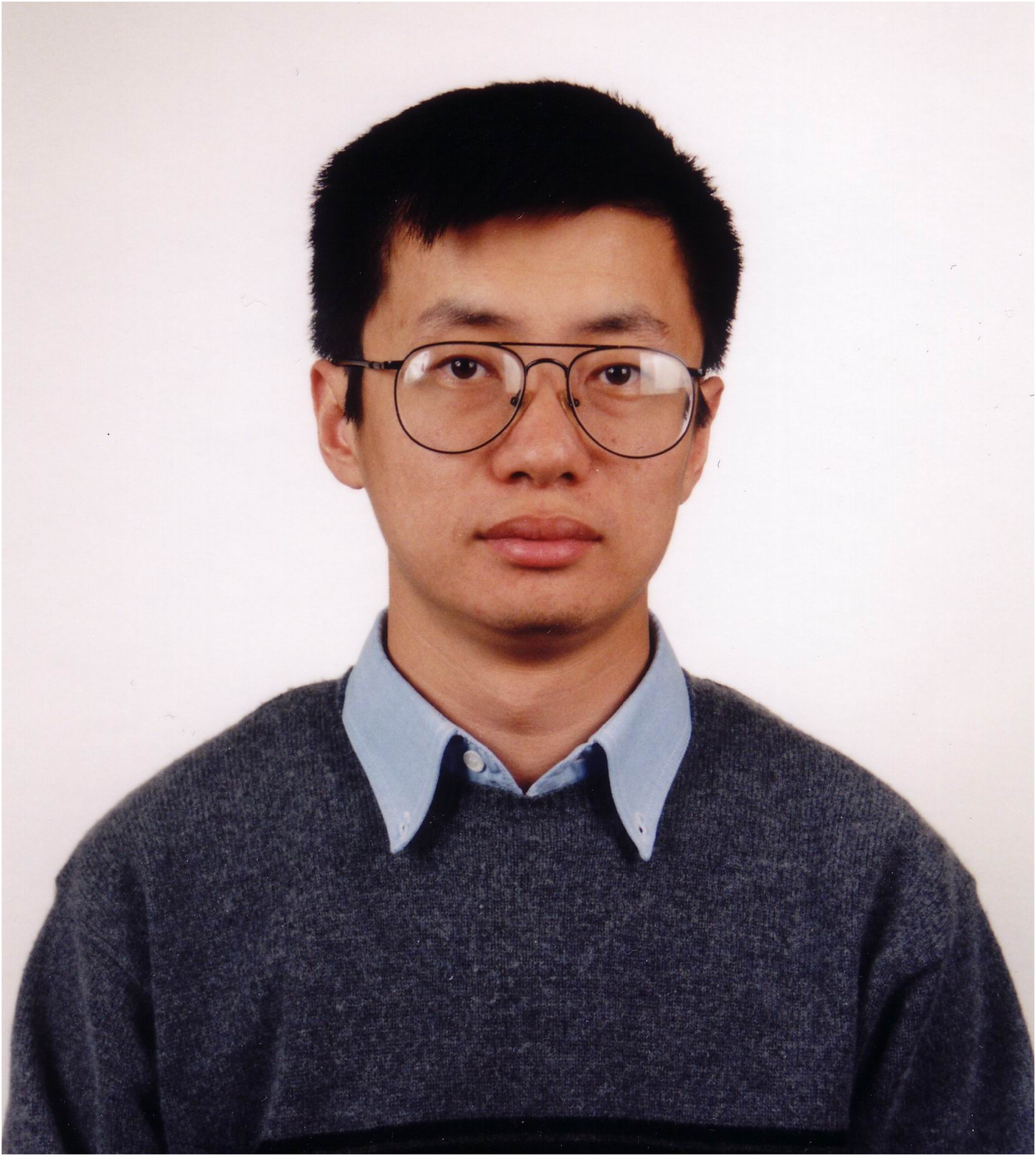}}] {Xiaodong Wang} (S'97-M'98-SM'04-F'08)
received
the Ph.D. degree in electrical engineering
from Princeton University, Princeton, NJ.

He is a Professor in the Department of Electrical
Engineering, Columbia University, New York.
His research interests fall in the general areas of
computing, signal processing, and communications,
and he has published extensively in these areas.
Among his publications is a book entitled Wireless
Communication Systems: Advanced Techniques
for Signal Reception (Prentice-Hall, 2003). His
current research interests include wireless communications, statistical signal
processing, and genomic signal processing.
Dr. Wang received the 1999 NSF CAREER Award, and the 2001 IEEE
Communications Society and Information Theory Society Joint Paper
Award. He has served as an Associate Editor for the IEEE TRANSACTIONS ON
COMMUNICATIONS, the IEEE TRANSACTIONS ON WIRELESS COMMUNICATIONS,
the IEEE TRANSACTIONS ON SIGNAL PROCESSING, and the IEEE TRANSACTIONS
ON INFORMATION THEORY.
\end{IEEEbiography}
\end{document}